\documentclass[10pt,a4paper]{article}
\usepackage[utf8]{inputenc}
\usepackage{geometry}
\usepackage{amsmath,amssymb}

\usepackage{tikz}
\usetikzlibrary{fpu, calc, arrows,decorations.pathmorphing,decorations.shapes,decorations.markings,backgrounds,positioning,fit,shapes.arrows,shapes.geometric,decorations.pathreplacing, shapes.symbols}

\usepackage{xcolor}
\usepackage[hidelinks]{hyperref}
\usepackage{authblk}
\usepackage{microtype}
\usepackage{subcaption}
\usepackage[aboveskip=1pt,labelfont=bf,labelsep=period,singlelinecheck=off]{caption}
\usepackage{graphicx}
\usepackage{colortbl}

\usepackage{tabularx}
\usepackage{multirow}

\usepackage{color}

\newcommand{\RR}{\ensuremath{\mathbb{R}}}
\newcommand{\ie}{}
\def\ie/{i.e.}
\newcommand{\fixme}[1]{\ignorespaces}
\newcommand{\fixedmaybe}[1]{\ignorespaces}
\newcommand{\fixedit}[1]{\ignorespaces}
\tikzset{twosimp/.style={fill opacity=0.6,fill=gray,draw opacity=0.9}}
\tikzset{twosimpred/.style={fill opacity=0.6,fill=red,draw opacity=0.9}}
\tikzset{threesimp/.style={fill opacity=0.8,fill=blue!60,draw opacity=0.9}}
\tikzset{belowdiag/.style={fill opacity=0.6,fill=gray,color=gray, draw opacity=0.6}}
\newcommand{\inputstr}{\ensuremath{\nu_{\text{ext}}/\nu_{\theta}}}

\newcommand{\mycite}[1]{\cite{#1}}
\newcommand{\myfootnote}[2]{\footnote{#2}}

\title{Topological exploration of\\ artificial neuronal network dynamics}

\author{Jean-Baptiste Bardin}
\author{Gard Spreemann}
\author {Kathryn Hess}
\affil{Laboratory for Topology and
  Neuroscience, Brain Mind Institute, École Polytechnique Fédérale de
  Lausanne, Lausanne, Switzerland}

\date{\today}

\begin{document}
\maketitle

\section*{Abstract}

One of the paramount challenges in neuroscience is to understand the dynamics of individual neurons and how they give rise to network dynamics when interconnected.
Historically, researchers have resorted to graph theory, statistics, and statistical mechanics to describe the spatiotemporal structure of such network dynamics. Our novel approach employs tools from algebraic topology to characterize the global properties of network structure and dynamics. 

We propose a method based on persistent homology to automatically classify network dynamics using topological features of spaces built from various spike-train distances. We investigate the efficacy of our method  by simulating activity in three small artificial neural networks with different sets of parameters, giving rise to dynamics that can be classified into four regimes. We then compute three measures of spike train similarity and use persistent homology to extract topological features that are fundamentally different from those used in traditional methods. Our results show that a machine learning classifier trained on these features can accurately predict the regime of the network it was trained on and also generalize to other networks that were not presented during training. Moreover, we demonstrate that using features extracted from multiple spike-train distances systematically improves the performance of our method.


\section{Introduction} 
A major objective in neuroscience is to understand how populations of interconnected neurons perform computations and process information. It is believed that the dynamics of a neuronal network are indicative of the computations it can perform. Its dynamics are affected by how the neurons are physically connected
and by the activity history of the neurons. Understanding this spatiotemporal organization of network dynamics is essential for developing a comprehensive view of brain information processing mechanisms, the \emph{functional connectome}. Two neurons can be considered ``functionally connected'' if their dynamics are similar or if one appears highly likely to spike causally after the other. The same notion of functional connectivity can be considered also on a macroscopic level, where one can study the causal relationships between brain regions. The notion can also be formalized for similarly structured systems from outside of neuroscience. Techniques like the one we present in this paper thus have broad applicability.

Furthermore, it is well known that certain neuronal systems can play multiple roles, characterized by different patterns of activity. For example, neurons in the thalamus have tonic or phasic behavior depending on the afferent signals and neuromodulators they receive or on different phases of the sleep cycle~\mycite{Weyand2011}. Another example is the hippocampus, which plays a role both in memory and in navigation. Researchers have also observed distinct rhythms in EEG recordings in awake and in sleeping rats~\mycite{BUZSAKI1990257}. 

An understanding of network dynamics is also of medical importance, as many neurological disorders are characterized by abnormal global or local activity. During epileptic seizures, for instance, EEG recordings show an increase in the amplitude of neural oscillations~\mycite{Fisher2005}. In Alzheimer's disease, one observes a shift in the power spectrum toward lower frequencies and a decrease in the coherence of fast rhythms~\mycite{Jeong2004}.

Partly because of the clinical importance of neural dynamics, various methods have already been developed to automatically detect abnormal regimes, for example those related to epileptic seizures~\mycite{Alkan2005, Tzallas2007, Khan2003}. The best ones rely on artificial neural networks. Here we propose a novel approach using techniques from topological data analysis, a part of applied mathematics.

Traditionally, neuroscientists have analyzed functional networks using pairwise neuron statistics and graph theory. Such methods often neglect certain global structures that may be present in the dynamics. The analysis of network dynamics using alternative methods from topological data analysis has recently enjoyed success~\mycite{Sapiro2008, curto2008cell, dabaghian2012, Giusti2015b, curto2017, Spreemann2018}. These methods provide information about connectedness, adjacency, and global structure such as holes (of various dimension) in a dataset\myfootnote{Space}{The dataset is turned into a mathematical \emph{space} in a manner detailed in the methods section. It is in the context of such a space that the notion of holes arises.}. In particular, persistent homology detects holes or cavities and quantifies how robust these are with respect to a threshold variable related to the dynamics of the system.

Several interesting properties of neuronal network structure and function have been revealed through these recent developments. For example, persistent homology has been applied to detect and characterize changes in the functional connectome in disorders such as autism and attention deficit disorder~\mycite{Lee2011}, in the Parkinson mouse model~\mycite{Im2016}, and after injection of a hallucinogenic substance~\mycite{Petri2014a}. It has also been employed to describe brain function during different tasks, such as multimodal and unimodal speech recognition~\mycite{Kim2015}.
Moreover, the homology of a digital reconstruction of the rat cortical microcircuit revealed that the brain substructure is non-random, that it is substantially different from various other null-models, and that its activity tends to concentrate in areas with greater local organization~\mycite{frontierspaper}. In the same article, homology was shown to distinguish the dynamics arising from different stimuli injected into the digital reconstruction.
It is also interesting to note that the mammalian brain seems to encode topological information, as there are strong indications that the place cells~\mycite{OKeefe1971} in the hippocampus build a topological, rather than geometric, representation of the animal's surroundings~\mycite{Dabaghian2014a}. The latter article also shows how such a map can be encoded by a spiking network of neurons.

To this day, the few articles in which persistent homology has been applied to \textit{in vivo} \mycite{Giusti2015b, Sapiro2008} or synthetic \mycite{Spreemann2018} spike data have all used spike train (Pearson) correlation as the distance between neurons. The use of correlations requires one to make specific assumptions about neural coding that may not be reasonable or relevant in all research areas. There exists a wide variety of spike train distances and similarity metrics, and the restrictiveness of their assumptions and the kind of information they encode can vary significantly. As we demonstrate in this paper, the appropriate notion of spike train distance to use depends on context, and it can also be beneficial to combine several of them.

In this paper, we simulate activity in an artificial network of neurons to generate spiking data and build weighted graphs based on various spike train similarities. These graphs are then transformed into topological spaces, which are analyzed using persistent homology. Finally, we extract simple features from topological invariants and use them to train a classifier for predicting the global network dynamics. These topological features are fundamentally different from those that might arise from graph-theoretic or other classical methods, as they take into account relations within triplets and not just pairs of neurons. The features are also different from other machine learning-suitable ones that have been suggested in the topological data analysis literature, such as persistence landscapes~\mycite{bubenik2015statistical}, persistence diagram heat kernels~\mycite{Reininghaus_2015_CVPR}, and persistence-weighted Gaussian kernels~\mycite{kusano2016persistence}.

Our results show that it is possible to perfectly predict network regimes from a few features extracted from the persistent homology. The trained classifier also predicts with high accuracy the dynamics of networks other than that on which it was trained. Finally, our results illustrate the importance of employing several spike train similarities, as the best performance was achieved using a combination of them.

\section{Results} 
In this section, we summarize our work, namely the simulation of network dynamics, the processing of spike trains, the topological analysis and feature selection, and the classification method, before presenting our results. A more detailed explanation of the method can be found in the methods section.

\subsection{Simulation of a downscaled Brunel network}
We consider simulated activity in the Brunel network~\mycite{Brunel2000}, a simple and well-studied \textit{in silico} model network of sparsely connected excitatory and inhibitory neurons. For computational reasons, we use a downscaled version of the network as described below.

\subsubsection{The Brunel network}
The Brunel network consists of two homogeneous sub-populations of excitatory (here indexed by $E$) and inhibitory (indexed by $I$) neurons modelled by a current-based leaky integrate-and-fire (LIF) model.

Each of the $N$ neurons has a membrane potential $V_{\text{m}}(t)$ whose dynamics are described by a differential equation. Once the membrane potential reaches a threshold value $V_{\theta}$, the neuron sends a spike through its synapses to its post-synaptic neurons, and its potential resets to a value $V_{\text{r}}$. The synapses are $\delta$-current synapses, i.e., after a delay $D$, each pre-synaptic spike induces  a positive (respectively, negative) jump in the membrane potential of the post-synaptic neurons if the pre-synaptic neuron is excitatory (respectively, inhibitory). The excitatory sub-population is four times larger than the inhibitory one ($N_{\text{E}} = 4N_{\text{I}}$) in accordance with cortical estimations~\mycite{Noback2007}, but their synapses are relatively weaker. Formally, if the excitatory synapses induce an increase of membrane potential of $J$, the inhibitory ones will induce a decrease of $-gJ$ for some $g>1$. Every neuron receives $K$ inputs coming from a fixed proportion $P$ of the neurons in each sub-population, \ie/, $K = P(N_{\text{E}}+N_{\text{I}})$. Furthermore, each neuron receives $C_E=P N_{\text{E}}$ external excitatory inputs from an independent Poisson population (of size $C_EN$) with fixed rate $\nu_{\text{ext}}$.  The relative synaptic efficiency ($g$) and the external population rate ($\nu_{\text{ext}}$) are the free parameters with respect to which we study network dynamics, once we have fixed the other model parameters, in particular $J$, $P$, and $D$. We adopt the convention of Brunel's original article~\mycite{Brunel2000} and express $\nu_{\text{ext}}$ as a multiple of the minimal external rate necessary to trigger spiking in the neurons without recurrent connections, denoted $\nu_{\theta}$.

Because computing persistent homology is expensive for large and dense spaces, which tend to arise from large and dense networks, the number $N$ of neurons  was reduced from $12{,}500$ in~\mycite{Brunel2000}  to $2{,}500$. Such a downscaling of the network while $N/K$ is kept constant will result in an increase in the correlation between the neurons~\mycite{VanAlbada2014}, more salient oscillations in the network dynamics~\mycite{Helias2013}, and potentially a loss in the diversity of network dynamics. To prevent these undesirable effects, a correction to the synaptic strength $J$ was applied, and the external population was modified according to \mycite{Helias2013}. Specifically, the synaptic strength $J$ was adjusted to keep $JK$ constant, and the rate of the external population $\nu_{\text{ext}}$ was increased. An external inhibitory population with appropriate rate was also introduced to preserve the mean and variance of the external inputs. The external rate correction is relevant only when neurons are expected to show irregular firing, \ie/, in the regimes where inhibition dominates ($g>4$).

We generated three versions of the Brunel network to validate our method across different networks.
\begin{description}
\item[Version 1:] Relatively sparse connectivity ($P=10\%$) and fast synaptic transmission ($D=1.5$ ms, $J = 0.1$ mV).
\item[Version 2:] Denser connectivity ($P=40\%$), fast and strong synaptic transmission ($D=1.5$ ms, $J = 0.2$ mV).
\item[Version 3:] Denser connectivity ($P=40\%$), slow and strong synaptic transmission ($D=3.0$ ms, $J = 0.2$ mV).
\end{description} 

Table~\ref{tab:brunelparams} gives a brief summary of the most important parameters. A complete description of the final model and parameter sets, following the formalism in the field~\mycite{Nordlie2009}, can be found in Tables~\ref{tab:brunel_model_1} and \ref{tab:brunel_model_2}.

\begin{table}[htbp]
\centering
\begin{tabular}{l|lll}
  \textbf{Symbol} & \textbf{Description} & \textbf{Value(s)} \\ \hline
  $N_{\text{E}}$ & \# excitatory neurons & $2000$ \\
  $N_{\text{I}}$ & \# inhibitory neurons & $500$ \\
  $C_{\text{E}}$ & \# exc.\ synapses per neuron & $200$–$800$ dep.\ on network ver. \\
  $C_{\text{I}}$ & \# inh.\ synapses per neuron & $50$–$200$ dep.\ on network ver. \\
  $J$           & Synaptic strength & $0.5$–$1.0$ $\text{mV}$ dep.\ on network ver. \\
  $D$           & Synaptic delay & $1.5$–$3.0$ $\text{ms}$ dep.\ on network ver. \\
  $g$           & Rel.\ synaptic efficiency & $\text{Free parameter}\in [2,8]$ \\
  $\inputstr$   & Rel.\ external rate & $\text{Free parameter}\in [1,4]$
\end{tabular}
\caption{Some of the more important parameters in the Brunel network. Tables~\ref{tab:brunel_model_1} and \ref{tab:brunel_model_2} give the full parameters of the model in the established formalism of the field.}
\label{tab:brunelparams}
\end{table}

\subsubsection{Simulations performed}
Each network was simulated for $20$ seconds of biological time with $28$ different values of the pairs of free parameters $g$ and $\inputstr$. These pairs form a rectangular grid in the parameter space, with $g$ taking values from $2$ to $8$ and $\inputstr$ taking values from $1$ to $4$. Since the network is connected according to a random model, each simulation was repeated $10$ times with different network instantiations, resulting in a total of $280$ simulations for each network version. The regimes of Brunel network dynamics are known to be robust~\mycite{Brunel2000}, which we also observed in our simulations. We were thus satisfied that $10$ instantiations suffice for our further analysis. We recorded the spiking times of all neurons, as well as the overall population firing rate, for all the simulations.

\begin{figure}[htbp]
  \centering
  \includegraphics[width=0.38\textwidth]{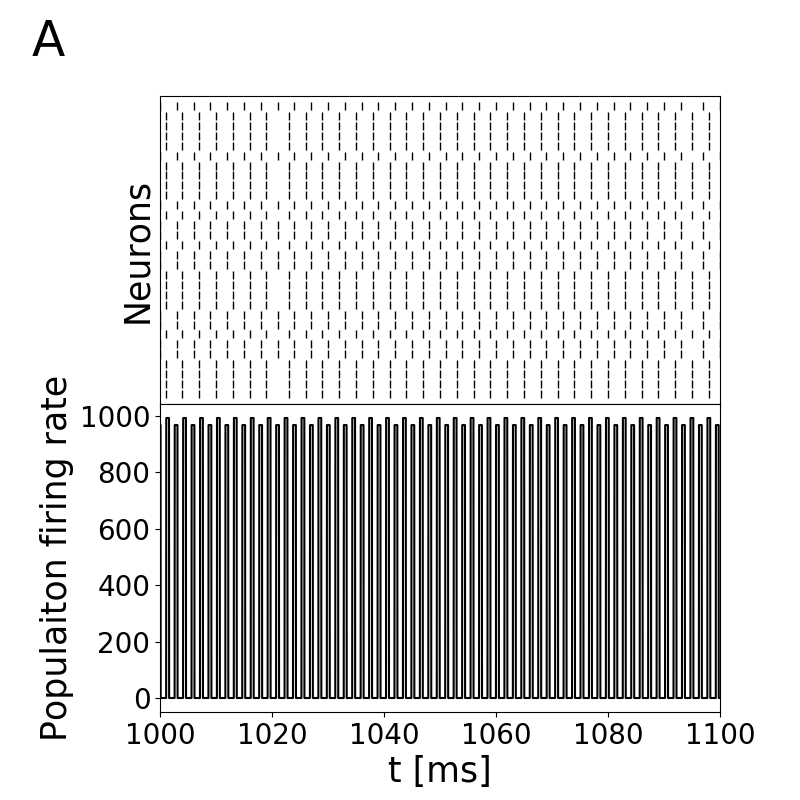}
  \includegraphics[width=0.38\textwidth]{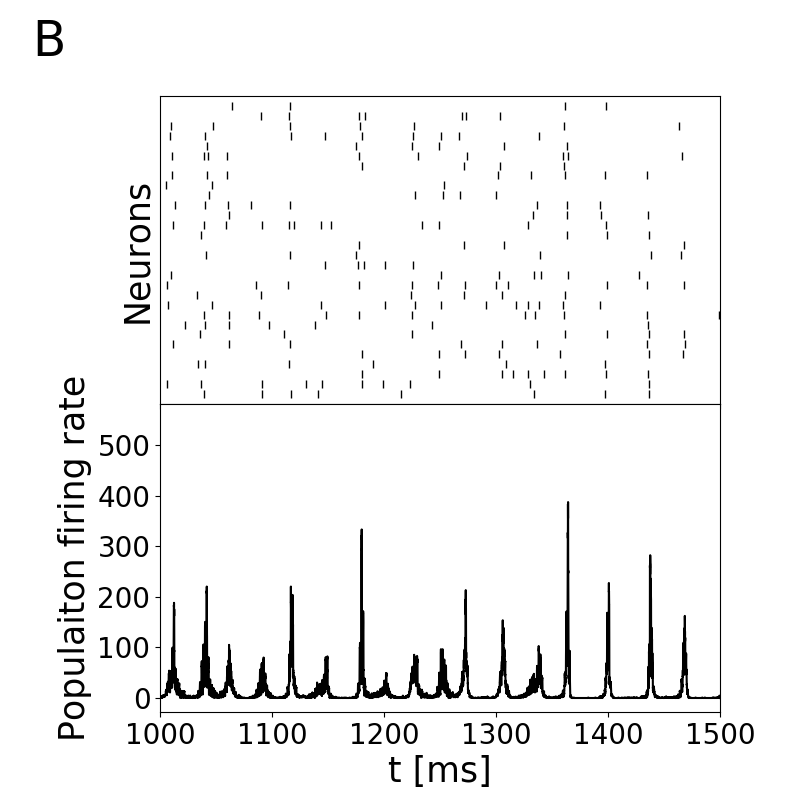}
  \includegraphics[width=0.38\textwidth]{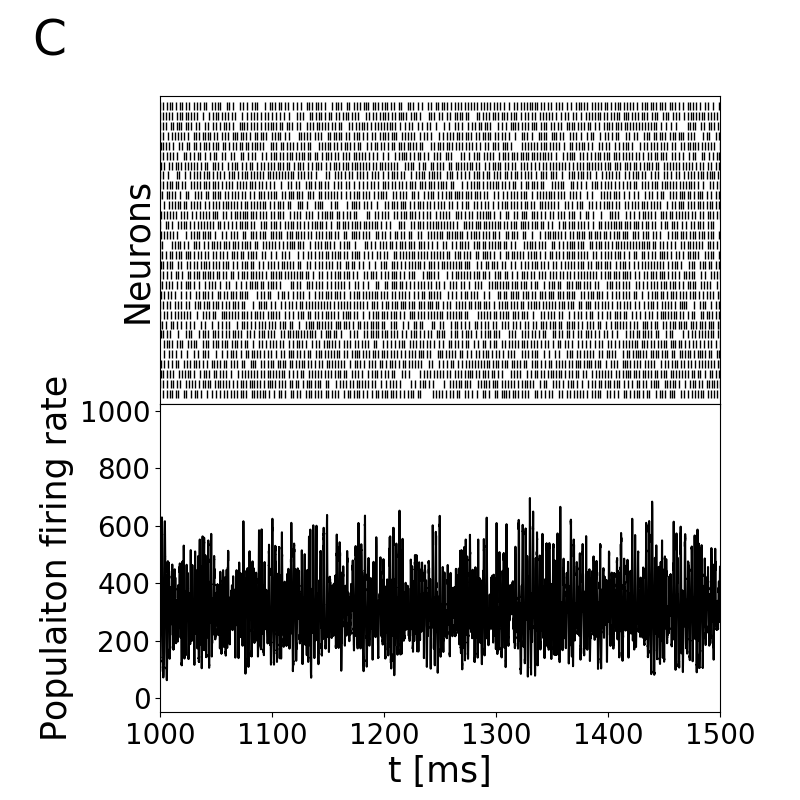}
  \includegraphics[width=0.38\textwidth]{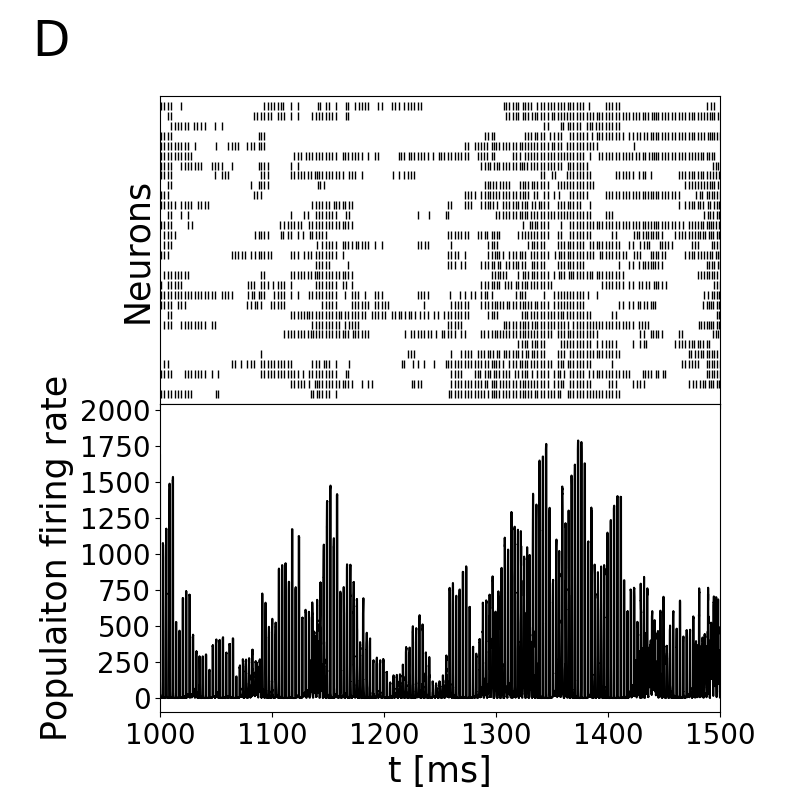}
  \caption
  {\small The four regimes produced by the downscaled Brunel network. Each panel shows the spiking activity (raster plot) of $30$ randomly selected neurons (top) and the population rate (bottom). \textbf{(A)} A $100$ ms segment of simulation in the synchronous regular regime (SR, $g = 2, \inputstr = 3$, version 1). \textbf{(B)} A $500$ ms segment of simulation in the synchronous irregular regime (SI, $g = 5, \inputstr = 1$, version 1). \textbf{(C)} A $500$ ms piece of simulation in the asynchronous irregular regime (AI, $g = 5, \inputstr = 2$, version 1). \textbf{(D)} A $500$ ms piece of simulation in the alternative regime (Alt, $g = 4, \inputstr = 2$, version 3).}
  \label{fig:simulations}
\end{figure}

Four distinct activity regimes, shown in Figure~\ref{fig:simulations}, were identified by manually inspecting the simulations and applying the same criteria as in~\mycite{Brunel2000}:
\begin{description}
\item[SR:] a regime characterized by synchronized neurons behaving as high-frequency oscillators, clustered in a few groups, similar to the synchronous regular regime (Figure~\ref{fig:simulations}A),
\item[SI:] a regime characterized by a slow oscillatory global pattern and synchronous irregular firing of individual neurons (Figure~\ref{fig:simulations}B),
\item[AI:] a regime characterized by asynchronous irregular firing of individual neurons (Figure~\ref{fig:simulations}C),
\item[Alt:] a regime characterized by neurons alternating between periods of silence and periods of rapid firing (Figure~\ref{fig:simulations}D). 
\end{description}

Note that the Alt\ regime is not present in the full-size network. This is not an issue for us, however, since our goal is to discriminate between different regimes, not to understand the Brunel network per se.

\begin{figure}[htbp]
  \centering
  \includegraphics[width=0.38\textwidth]{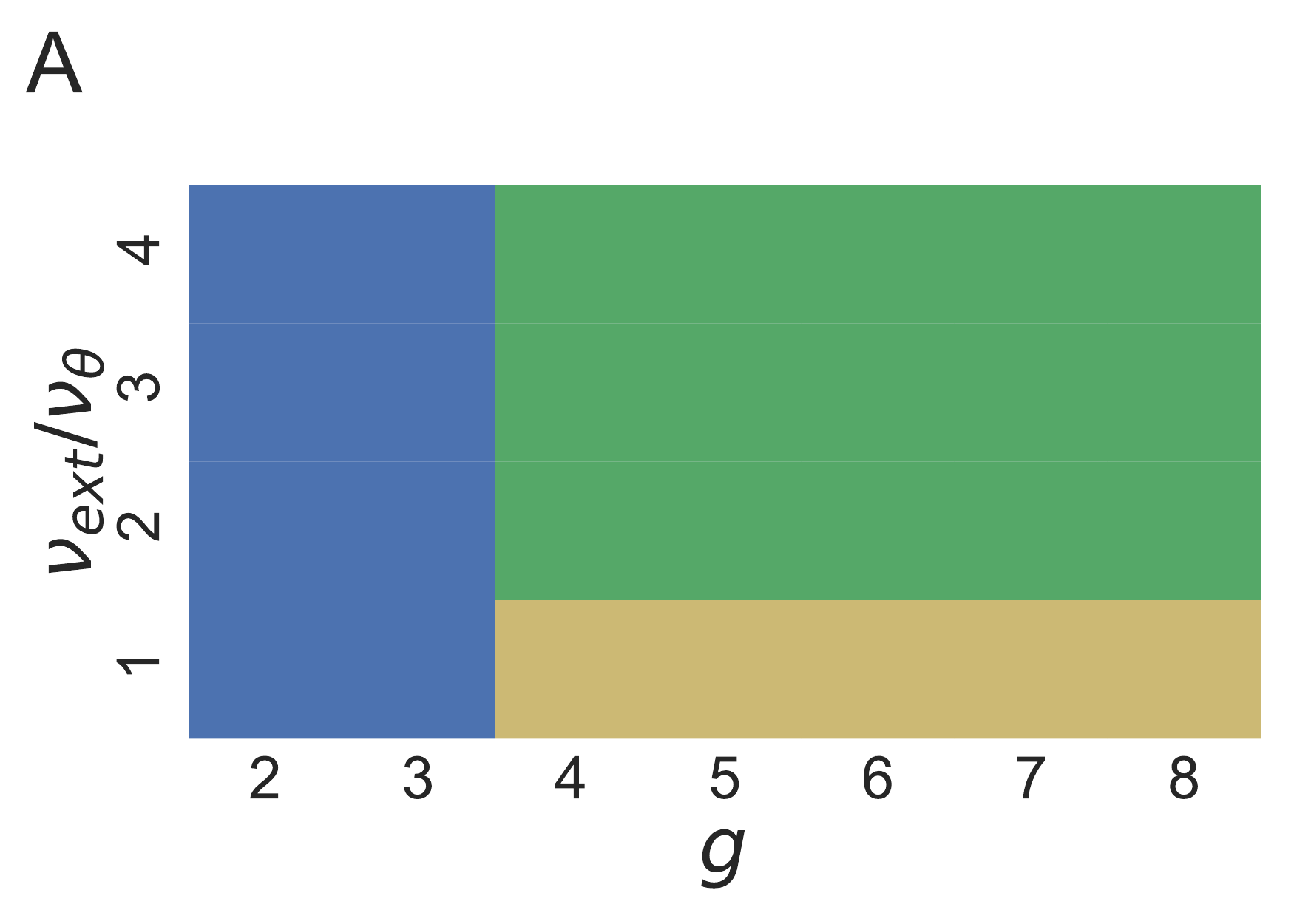}
  \includegraphics[width=0.38\textwidth]{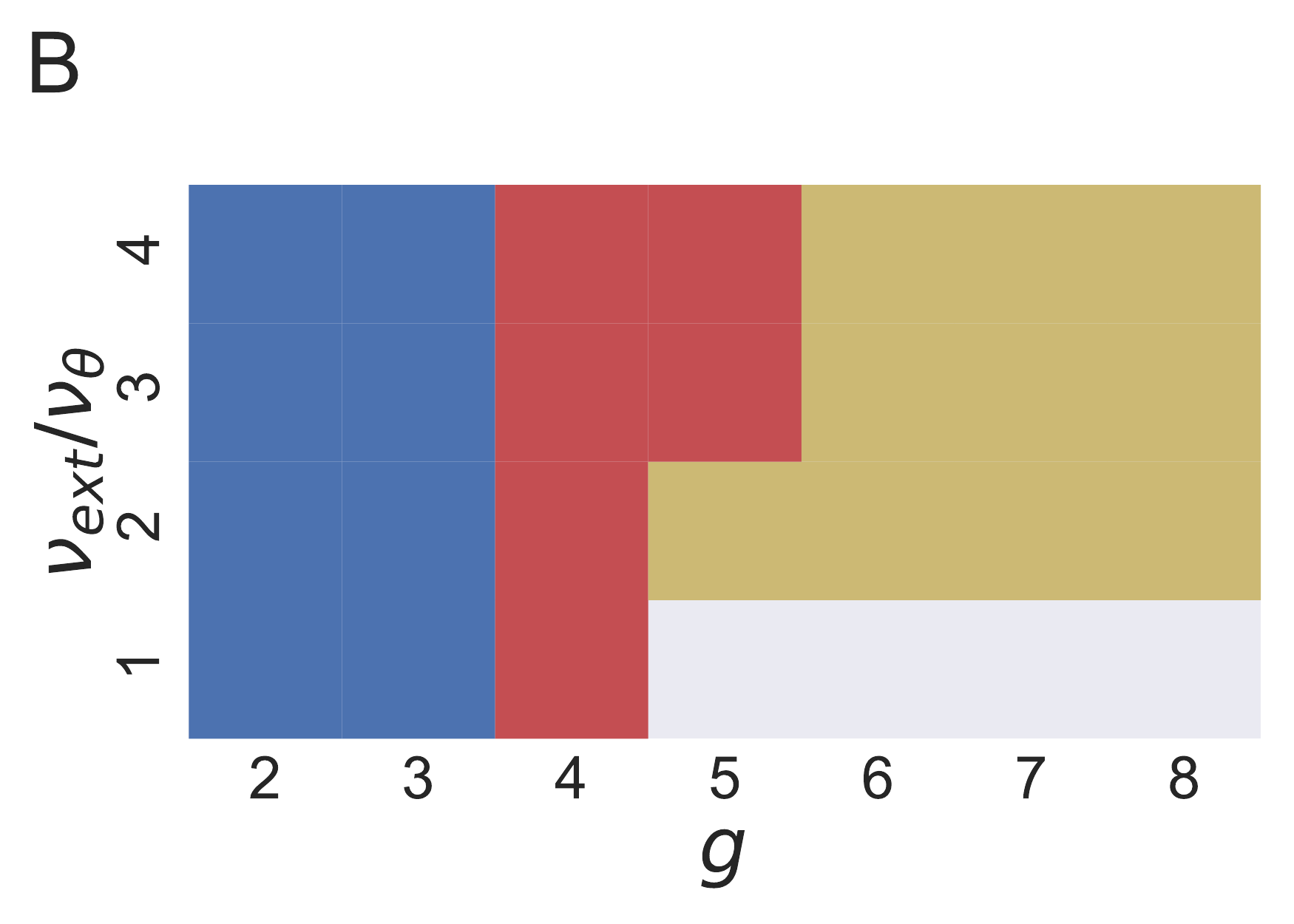}
  \includegraphics[width=0.38\textwidth]{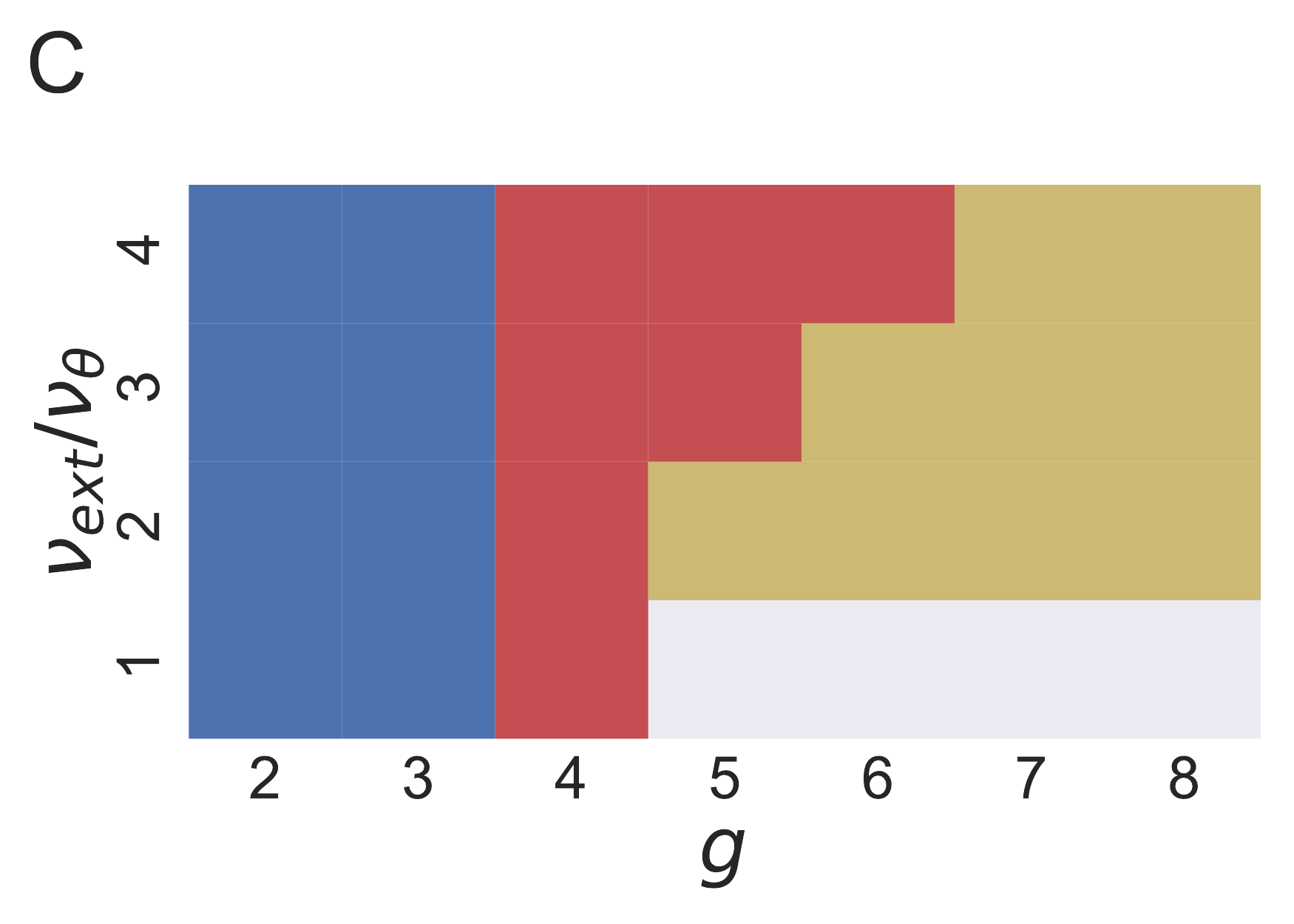}
  \includegraphics[width=0.15\textwidth]{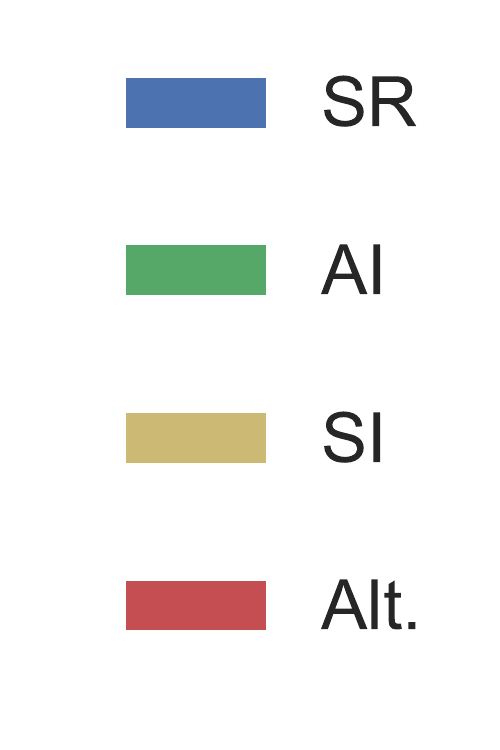}
  \caption
  {\small Diagrams of the different regimes for each version of the Brunel network (\textbf{(A)}, \textbf{(B)} and \textbf{(C)} corresponding to versions 1, 2, and 3 of the network, respectively) in the parameter space $(\inputstr, g)$. The white areas indicate where no neurons fired and no data is available.} 
  \label{fig:PD}
\end{figure}

For each of the three networks, we visually identified the network regime for every pair of parameters $(g, \inputstr)$ by the same criteria as used in~\mycite{Brunel2000}. We believe that no automated method exists to verify the criteria set out in that paper, but that this visual identification is quite consistent. The result is shown in Figure~\ref{fig:PD}. The simulations in which none of the neurons fired were removed from the analysis ($40$ simulations for versions 2 and 3). Note that the first network (version\ 1) does not exhibit the Alt regime,  while versions 2 and 3 do not exhibit the AI regime. This issue is addressed in the methods section concerned with machine learning.

\subsection{Spike train similarities}
We used three different measures of spike train similarity to compare the recorded neuron activity in the networks.

One is the widely used Pearson correlation. It is often employed in analyzing spiking data because it has been shown to encode particular information that is not present in the firing rate alone. For example, in the auditory cortex of the marmoset, Pearson correlation encodes the purity of sounds~\mycite{deCharms1996}. It can also be used to infer connectivity or extract information about  network function~\mycite{Cohen2011}. However, it is tied to the correlation population coding hypothesis~\mycite{Panzeri2015}, and thus may not be relevant to the problem at hand. We therefore also employed two complementary measures: SPIKE-synchronicity~\mycite{Kreuz2013} and SPIKE-distance~\mycite{Kreuz2015a}. Both are exploratory measures relying on an adaptive time window to detect cofired spikes and involve a pairwise similarity measure of spike trains. Conceptually, the size of the window depends on the local firing rate of the two neurons under consideration. If one of the neurons has a high local firing rate, then the time window will be short, while if both neurons have low local firing rates, the time window will be longer. SPIKE-synchronicity is the fraction of cofired spikes according to this adaptive window, while SPIKE-distance is the average over time of a dissimilarity profile between spike trains. 
See equations (7)--(13) in \mycite{Kreuz2015a} for details.

\begin{figure}[htbp]
  \centering
  \includegraphics[width=0.6\textwidth]{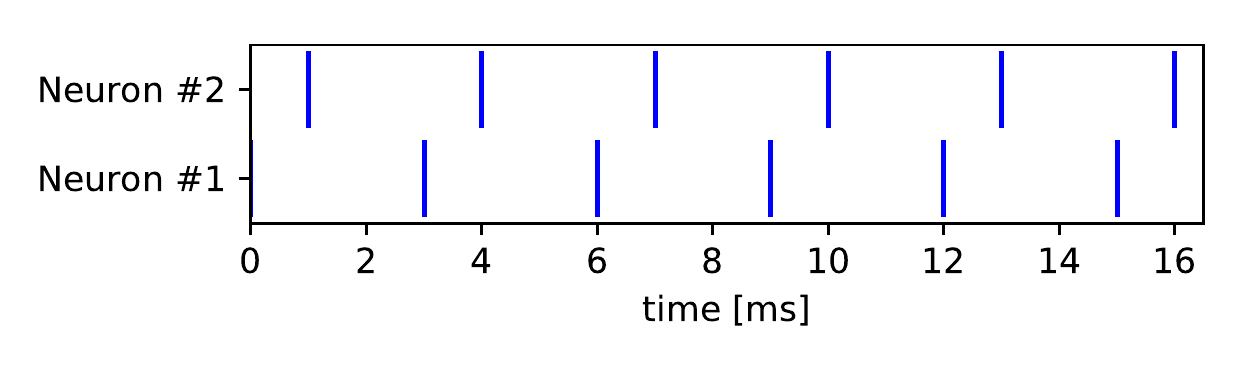}
  \caption{A pair of spike trains for which Pearson correlation is $0.21$, SPIKE-synchronicity is $1.00$ and SPIKE-distance is $0.30$.}
  \label{fig:small_spike_trains}
\end{figure}

Figure~\ref{fig:small_spike_trains} shows an example of two very regular spike trains for which the three measures encode very different relationships.

We computed Pearson correlations by time-binning the spike trains with a $2$ ms time window and a binning error correction, as described in the methods section. The SPIKE measures were computed using the Python package PySpike~\mycite{Mulansky2016}. 

\subsection{Persistent homology}
In \emph{topology}, a branch of mathematics, one works with very general objects called \emph{spaces}. Spaces have a notion of ``nearness'', but in general lack more geometric structure such distances or angles, as well as familiar algebraic structure. The unaccustomed reader may still gain intuition about what is meant by a space by thinking of geometric objects. One should still keep in mind that the spaces we consider may be defined entirely intrinsically without any reference to some ambient Euclidean space.

\emph{Algebraic} topology, then, concerns itself with describing a space $X$ in terms of algebraic invariants that capture global properties of the space. Examples include the Betti numbers, which can be thought of as the numbers of components and of $n$-dimensional unfilled cavities in the space. As these notions can be defined intrinsically, \ie/, without any reference to how the space is embedded in Euclidean space, they are useful in analyzing spaces arising from abstract data where no such embedding can be constructed in a principled way. We consider here only the zeroth Betti number $b_0(X)$, which is the number of connected components, and the first Betti number $b_1(X)$, which is the number of one-dimensional unfilled loops. In the special case of graphs --- which is \emph{not} the case we consider --- these are precisely the number of connected components and the number of cycles, respectively.

The spaces we study here will be built from spiking data, and we are interested in how these algebraic invariants change as a function of a spike train similarity threshold. We therefore build a \emph{filtration}, a multi-scale sequence of spaces depending on a threshold, and compute \emph{persistent homology}, a multi-scale invariant that captures Betti numbers (which are then often also referred to as \emph{Betti curves} to reflect their scale-/threshold-dependent nature). For details, see the methods section devoted to the topological analysis. More background information can be found in the survey~\mycite{ghrist2008barcodes} and its bibliography. 

As is common in topological data analysis, we follow the convention that edges with low weights are to be considered ``most important'' and enter first in the filtration (see the methods section for the definition). The correlation and SPIKE-synchronicity values were transformed through the function $x\mapsto 1-x$ so that they range from $0$ to $1$ with $0$ being the value assigned to a pair of identical spike trains (i.e., we work with dissimilarity measures).

\subsection{Classifying network dynamics} 
We used the output of persistent homology of the space built from pairwise spike train similarities as an input feature for machine learning in order to discern information about the global network dynamics. While we did not investigate the matter, it could be that even simpler features --- such as simplex counts across filtrations, or other features based on just counting local properties --- suffice. Such features rely less on the relationships between neurons, and are sensitive only to the creation (or not) of individual simplices, without considering how these simplices build greater structures.

From every filtration, four simple features were extracted:
\begin{itemize}
\item from the Betti-$0$ curve, the area under the curve and the filtration value at which it starts to decrease,
\item from the Betti-$1$ curve, the global maximum and the area under the curve.
\end{itemize}
As a result, a total of $12$ features were extracted from each simulation (four features per filtration, three filtrations from the three spike train similarity measures). For some simulations in the SR regime, all the pairwise similarities attained the maximal value, resulting in a space with no topological features and a constantly zero Betti-$1$ curve. The filtration value at which the curve starts to decrease was defined to be $0$ in this case.

\begin{figure}[htbp]
  \centering
  \includegraphics[width=\textwidth]{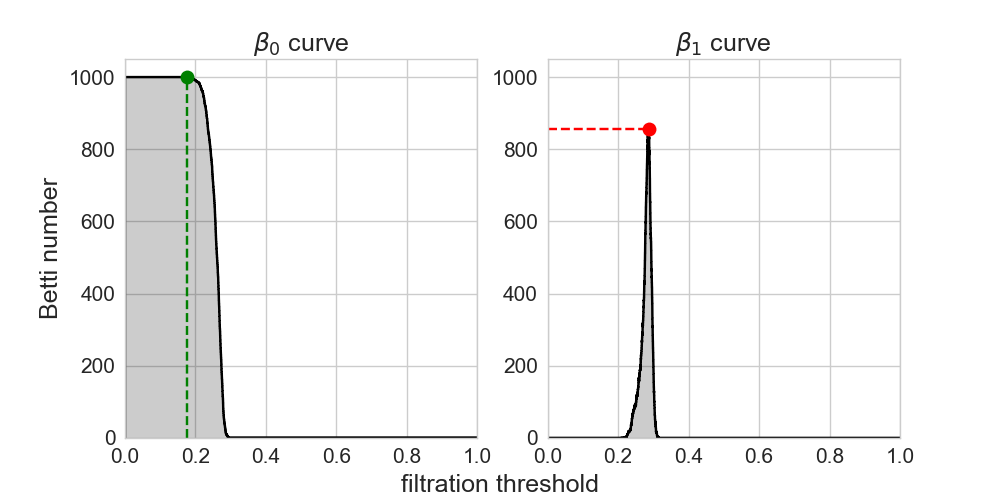}
  \caption[]
  {\small Example of three of the features extracted from a filtration: the filtration threshold at which the Betti-0 curve (left) starts to decrease, the global maximum (red) of the Betti-1 curve, and the area under each curve (grey area)}
  \label{fig:features}
\end{figure}

Before doing any classification, potentially good features were selected by plotting all the features against each other for the samples coming from network version $1$. Six features were selected by visual inspection because they were deemed to produce non-overlapping clusters. These features are the area under the Betti-$0$ curve for the three similarity measures, the area under the Betti-$1$ curve for correlation and SPIKE-synchronicity, and the maximum of the Betti-$1$ curve for the SPIKE-distance. These features were among the ones with the highest mutual information in the three networks, although their ranking varied between the networks. A section under methods is devoted to the feature selection process; see in particular Figure~\ref{fig:MI_2}.

Four different training sets were used for classification, three of which were composed of randomly selected samples ($90\%$) coming from a specific network version, while the last set was the union of the three other sets. For each training set, an $L^2$-regularized support vector machine (SVM) classifier was trained to identify the different regimes. The classifier was composed of four sub-classifiers, each of which had to distinguish one particular regime from the others. The final decision was computed in a one-vs-rest manner. The regularizing hyperparameter was selected with a $10$-fold cross-validation.

When assessing the performance of the classifier for network version $k$ using multi-class accuracy, we validated it on three test sets: one composed of the $10\%$ of the samples from version $k$ that had not been used for training and two containing all the valid samples from one of the other network versions. A sample was considered valid if it was labeled with a regime that was present in the training sets. For example, when the classifier trained on version $1$ was tested on version $2$, the samples labeled Alt\ were ignored, since no version $1$ networks exhibit the Alt\ behavior. The performance accuracy  and the numbers of valid samples are reported in Table~\ref{tab:ML_res_2}. The trained classifiers all achieved perfect accuracy ($100\%$) on the network version they were trained on, indicating that the topological features extracted are sufficient to perfectly discriminate the regimes of the training network. Moreover, they also generalized well to other versions, with $94.26\%$ accuracy on average, suggesting that the topological features extracted are consistent across the three network versions.

Additionally, we combined all the samples from the three networks and used a $90\%$-$10\%$ training-testing sample repartition, attaining perfect classification (``All versions'' row in Table~\ref{tab:ML_res_2}) of the four regimes. This provides the complementary information that the Alt and AI regimes are also distinguishable from one another, since none of the network versions can exhibit both regimes.


\begin{table}[htbp]
\centering
\begin{tabular}{l|llll}
                          & \multicolumn{4}{c}{\textbf{Testing set}}  \\
\textbf{Training set}     & Ver.\ 1    & Ver.\ 2    & Ver.\ 3   & All ver. \\ \hline
Version $1$                    & $100\%$ ($28$)   & $86.67\%$ ($180$) & $91.18\%$ ($170$) & $89.68\%$ ($378$) \\
Version $2$                  & $97.69\%$ ($130$) & $100\%$ ($24$)   & $93.33\%$ ($240$) & $95.18\%$ ($394$)\\
Version $3$                    & $99.23\%$ ($130$)& $99.17\%$ ($240$)& $100\%$ ($24$)  & $99.23\%$ ($394$) \\
All versions                  & $100\%$ ($28$)  & $100\%$ ($24$) & $100\%$ ($24$)  & $100\%$ ($76$)  
\end{tabular}
\caption{
Classification accuracy for each pair of training and testing sets. The training sets were comprised of $90\%$ of the samples coming from a specific network (version 1, 2 or 3), or $90\%$ of all samples (all versions). The testing sets contained the remaining samples from the network version that are not used for training. i.e., the remaining $10\%$. The number of samples in every testing set is reported in parentheses. 
}
\label{tab:ML_res_2}
\end{table}

\begin{figure}[htbp]
  \centering
  \includegraphics[width=0.4\textwidth]{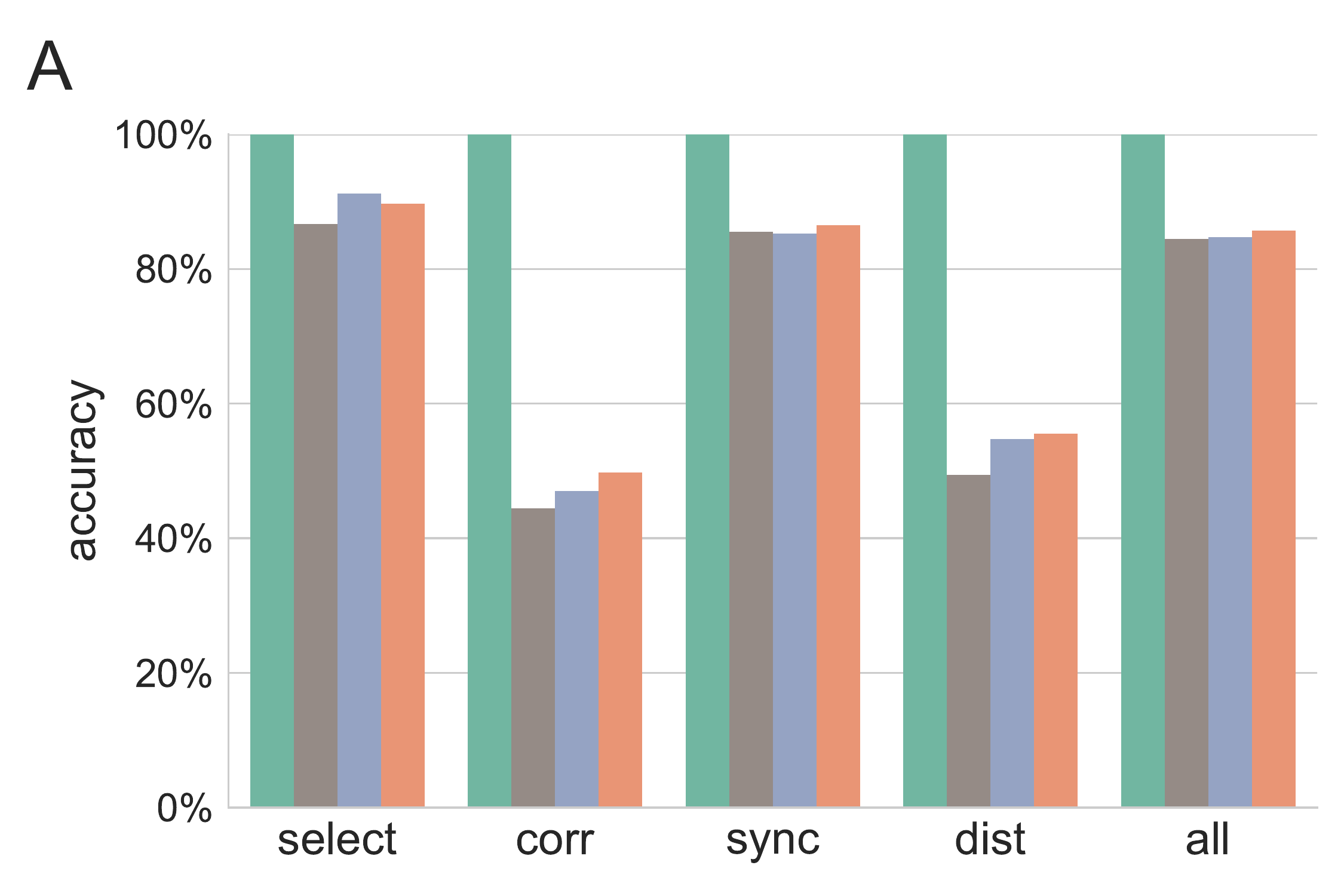}
  \includegraphics[width=0.4\textwidth]{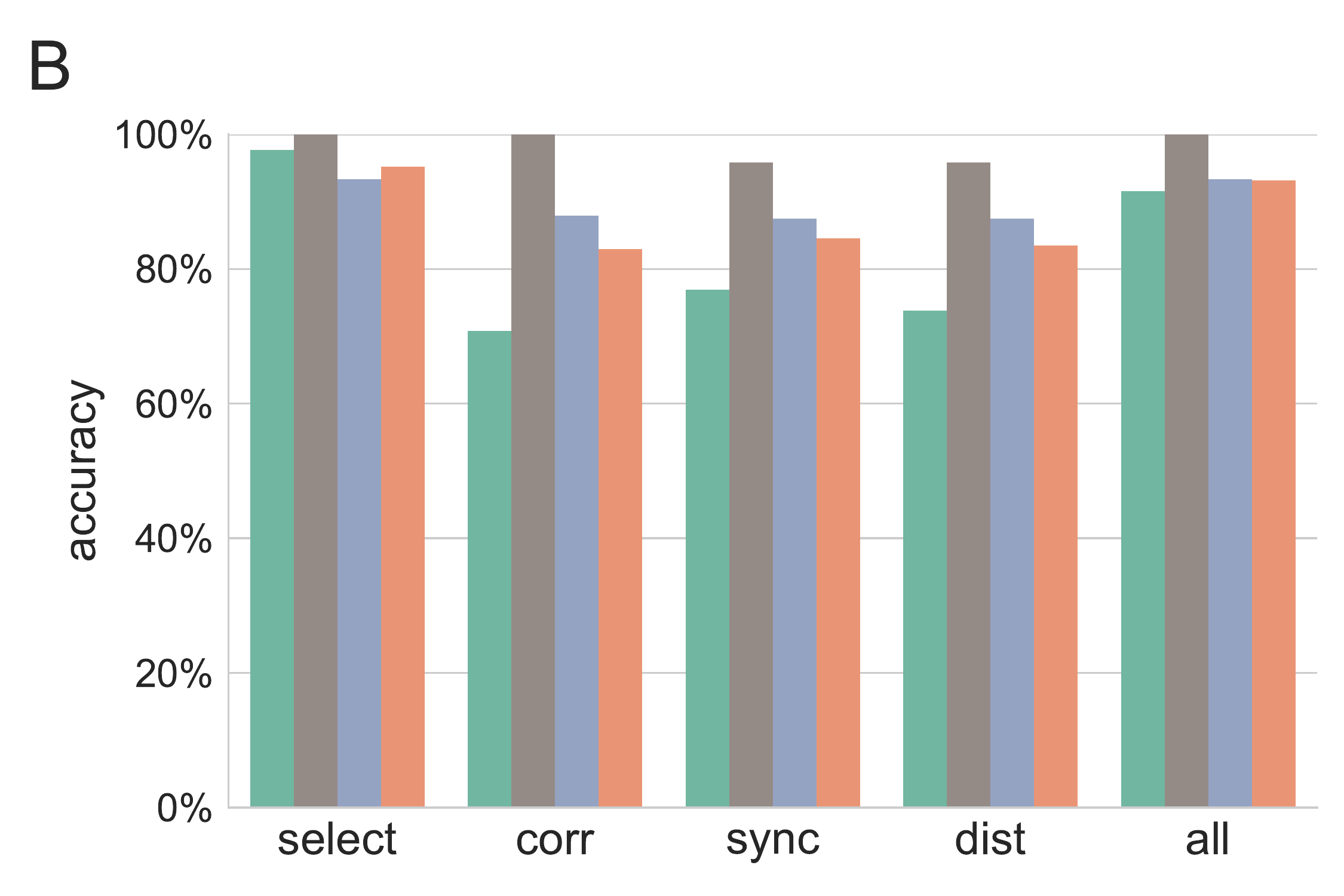}
  \includegraphics[width=0.4\textwidth]{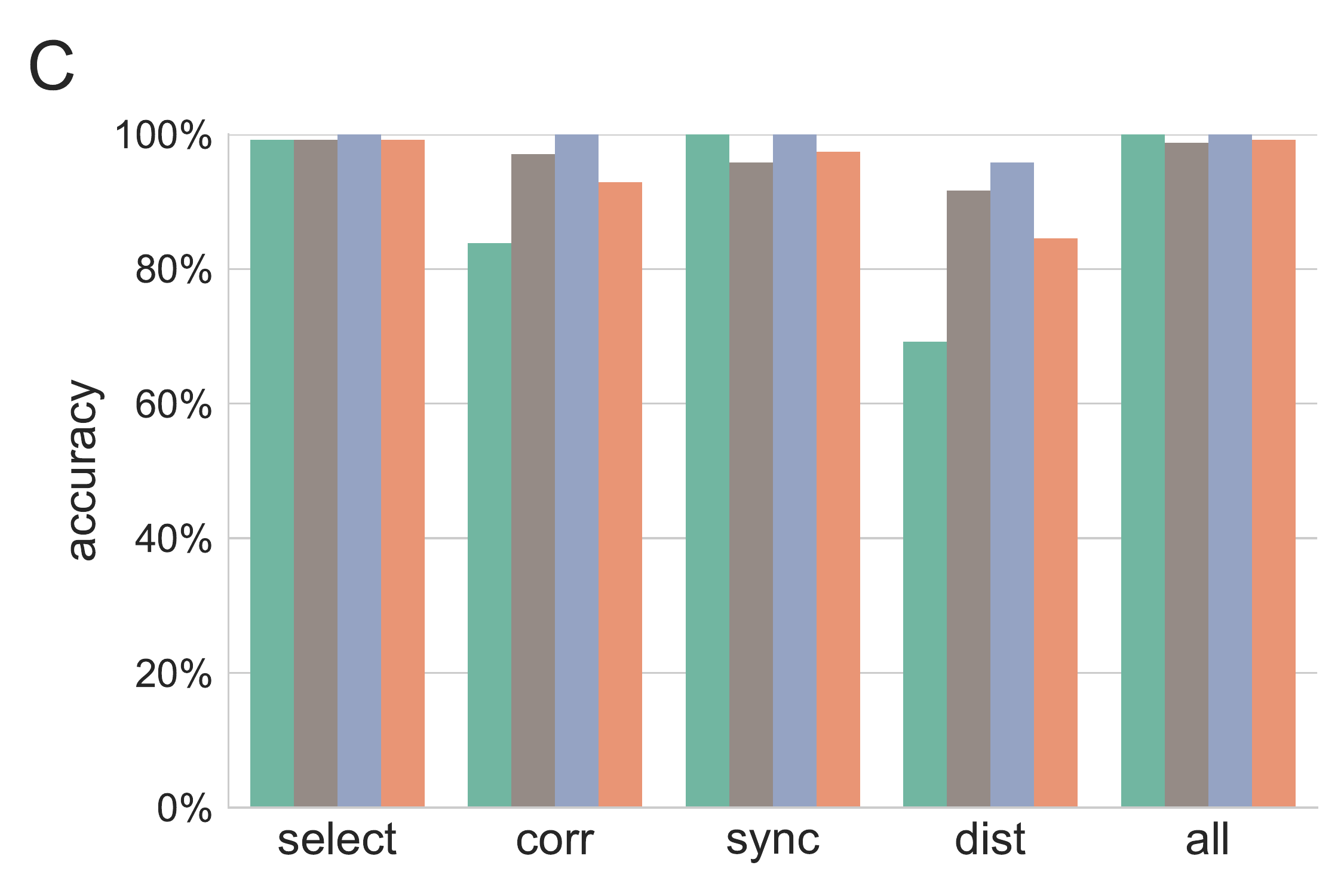}
  \includegraphics[width=0.4\textwidth]{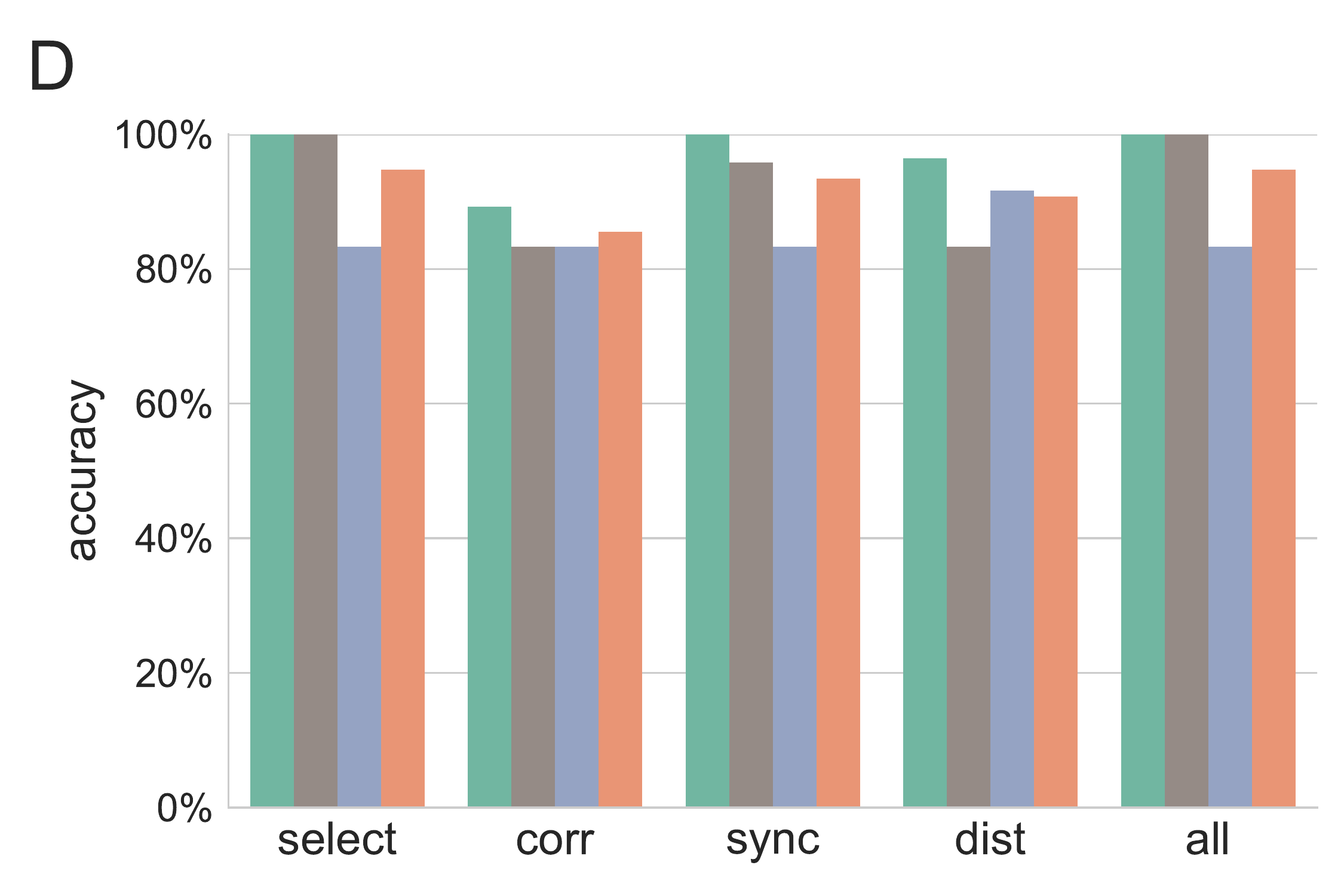}
  \includegraphics[width=0.1\textwidth]{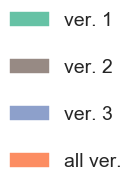}
  \caption[]
  {\small Testing accuracy of the classifier trained on samples from version $1$ \textbf{(A)}, version $2$ \textbf{(B)}, version $3$ \textbf{(C)} and all the samples \textbf{(D)}. In each panel, the classification accuracy for the test samples from each network and all the samples together is reported for five sets of features. ``Select'' designates the features we visually selected. ``Corr'', ``sync'' and ``dist'' designate the features extracted using the correlation, SPIKE-synchronicity, and SPIKE-distance, respectively. ``All'' designates the set of all the features. \fixedit{G: I'm not a huge fan of line plots when there's no reasonable notion of what the stuff between the data points means. Could you maybe replot this as a bar plot, JB? JB: is it better now? G: Yep, great!}}
  \label{fig:acc}
\end{figure}

Finally, we checked whether the persistent homology-derived features provide complementary information when based on different similarity measures, and thus whether it can be advantageous to use several of them together for classification. The same classification experiments were repeated using either all the computed features or only the features coming from one of the similarity measures. The classifier accuracies were compared with those from the previous computations (Figure~\ref{fig:acc}). Although the accuracies obtained using features coming from a single similarity measure were satisfactory (on average $79.10\%$, $92.35\%$, and $80.73\%$ accuracy for correlation, SPIKE-synchronicity, and SPIKE-distance, respectively), better performance was consistently attained by a combination of measures. This is consistent with our expectations, since the similarity measures we use give significantly different orderings of the pairs of spike trains. Moreover, selection of potentially good features yields the best results ($94.94\%$ and $96.63\%$ average accuracy with the ``all'' set and ``select'' set, respectively).

\section{Discussion} 
In this paper we analyzed the dynamics of spiking networks of neurons using features derived from persistent homology. We generated three versions of a simple artificial network of LIF neurons (a downscaled version of the Brunel network) by modifying connectivity density, synaptic delay, and synaptic strength. Activity in the networks was then simulated with $28$ pairs of the free parameters (external population rate and relative synaptic efficiency values). Across all the simulations, four regimes of activity were observed based on the pattern of the global population firing rate and the individual neuron spiking times. 
 
For each simulation, we computed three pairwise spike train similarity measures: Pearson correlation, SPIKE-synchronicity, and SPIKE-distance. We computed the persistent homology of the flag complex of the weighted graph coming from each similarity measure and extracted simple features from the zeroth and first Betti curves. The interesting features were selected by visual inspection of the sample distribution. Finally, an SVM classifier was trained to identify the dynamics regimes of the simulations.

Our experiments showed that it is possible to perfectly predict the dynamic regime in simulations coming from the network trained on, and from other networks with a high degree of accuracy, as long as some samples of the regimes in question were available during training. 

We also illustrated the importance of using and combining several similarity measures. Indeed, SPIKE-synchronicity carries more information, and does so more consistently across the network versions, than the other two measures, but the best accuracies were consistently obtained when an ensemble of features selected by visual inspection was used. Moreover, if one were to automatically select the features based on a score, we showed that the mutual information between features and the regime label is a good indicator to consider. 

We tested our method in the context of a simple network. It would be interesting to test it also with more complex networks, with neurons and synapses modeled in greater detail. Topological features can also be extracted from other types of neural data, such as the population firing rate or neuron voltage traces. We consider the examination of how the topological methods perform in classifying such data as interesting future work.

The present paper does not discuss unsupervised methods. We did perform small exploratory experiments in which the persistent homologies arising from the spike trains were transformed into real-valued functions by means of the persistence heat kernel~\mycite{Reininghaus_2015_CVPR}. These functions were then considered as points in a metric space of functions, and embedded in a low-dimensional Euclidean space using multi-dimensional scaling. While we ultimately failed at satisfactorily clustering these Euclidean points in an unsupervised way, it is an approach that may be worth considering in future work. An unsupervised version of our method may be useful for real-time detection of previously unknown ephemeral regimes of dynamics.

As mentioned, our topological features are simple, but perhaps not the simplest possible. It would be interesting to see a comparison between the performance of our features and features that are merely summaries of local properties, such as the total count of simplices of various dimensions as a function of the filtration parameter. We suspect that such features pick up too little of the organizational structure \emph{between simplices}, and will thus decrease the classifier's performance without any significant reduction in computation time.

Here we have illustrated just one concrete use of topological data analysis (TDA) in the study of network dynamics, but the class of methods should be applicable to a wide variety of systems from within and without neuroscience. To the best of our knowledge, there have been no previous attempts at applying TDA to automatic detection of regimes in spiking neural networks, since they are usually identified analytically~\mycite{Brunel2000,Helias2013} and can often be discriminated visually. However, a topological approach to this task may be interesting in recordings of real data, such as  EEG or fMRI. One might, for example, investigate the feasibility of solving a more subtle task, such as automatic detection of movement intention or seizure detection in epileptic patients.

Although great progress has been made in neuroscience since the first recording of a neuron activity in 1928 \mycite{Adrian1928}, a unified model of the brain across its different scales is still lacking, and many hard challenges have barely been attempted. Recent work has highlighted how TDA could help shed new light on both brain structure and dynamics and are promising advances towards a more comprehensive understanding of the brain. The method we have outlined in this paper takes a novel view of one challenge, the automated classification of neuronal dynamics, by considering features that are topological in nature. We believe that including such features will be of great help in the understanding of both structural and dynamical aspects of neuronal networks and other similarly structured systems. 

While we in this work considered only a very specific system, namely simulated spiking neurons, our method should be applicable in a wide variety of settings, both inside and outside of neuroscience. At its core, the method just requires ways of comparing timeseries, and it may therefore be useful in classifying regimes in general dynamical systems (perhaps coupled with, or complementing, delay embedding-based methods~\mycite{perea2015sliding} in the case of smooth timeseries). Within neuroscience, EEG recordings and fMRI BOLD signals immediately suggest themselves as data that can be studied in this way.

\section{Methods}

We give here the full details of our computations, and expand on the topological constructions involved in the analysis.

\subsection{Network simulations}
A complete specification of the three simulated networks following formalism and notation commonly used in the field~\mycite{Nordlie2009} can be found in Tables~\ref{tab:brunel_model_1} and~\ref{tab:brunel_model_2}.

All the networks were simulated with $28$ pairs of parameter values for the relative strength between inhibition and excitation $g$ (integer values from $2$ to $8$) and the external population rate $\inputstr$ (integer values from $1$ to $4$). The systems were simulated $10$ times for each parameter pair, for a total of $280$ simulations per network. The simulations were performed with the Brian2 simulator~\mycite{Brian}, with a time step of $0.01$ ms and a total biological time of $20$ seconds.
Because of the downscaling of the network, the synaptic transmission $J$ was increased compared to that used in~\mycite{Brunel2000} in order to keep $C_E J$ constant, and an external inhibitory population was introduced~\mycite{Helias2013} when the spiking of neurons was expected to be irregular~\mycite{Brunel2000}. This external population was modeled by a Poisson process with rate
\begin{equation}
\label{eq:rate_bal}
\nu_{\text{bal}} = \frac{\sigma_{\text{i}}^2 - \sigma_{\text{loc}}^2 - \tau_m \nu_{\text{ext}} J_{\text{ext}}^2}{\tau_m J_{\text{ext}}^2 (1+g^2)}
\end{equation}
as in~\mycite{Helias2013}. Here, $\sigma_{\text{i}}$ is the variance of input in the original network and $\sigma_{\text{loc}}$ is the variance due to local input from recurrent connections in the downscaled network. The variances can be approximated as
\begin{align}
\sigma_{\text{i}}^2 = C_E^{\ast} {J^{\ast}}^2 \tau_m \left( \nu_{\text{ext}} + \nu_{0}  \left( 1+g^2 \gamma  \right)  \right) \label{eq:sigma_i} \\
\sigma_{\text{loc}}^2 = C_E J^2 \tau_m \nu_{0}  \left( 1+g^2 \gamma  \right). \label{eq:sigma_loc}
\end{align}

The parameters for the original network are labelled by an asterisk and differ from their counterpart in the downscaled network by a scaling factor $\alpha$ such that $C_E = C_E^{\ast}/\alpha$ and $J = \alpha J^{\ast}$. From equations~(\ref{eq:sigma_i}), (\ref{eq:sigma_loc}) and (\ref{eq:rate_bal}), we obtain 
\begin{equation}
\nu_{\text{bal}} = \frac{\left( 1/\alpha - \alpha \right) \nu_{\text{ext}} + \left( 1/\alpha - 1 \right) \nu_0 \left( 1+g^2 \gamma  \right) }{\tau_m J_{\text{ext}}^2 (1+g^2)}.\label{eq:final}
\end{equation}
Here $\nu_{0}$ is the stationary frequency of the original population and can be approximated by~\mycite{Brunel2000} 
\begin{equation}
\nu_{0} = \frac{\nu_{\text{ext}} - \nu_{\theta}}{g\gamma - 1}.
\end{equation}

\begin{table}[htbp]
\centering
\begin{tabularx}{\textwidth}{lX}
\hline
\rowcolor[HTML]{333333}
\color[HTML]{FFFFFF} A                       & \multicolumn{1}{c}{{\color[HTML]{FFFFFF} Model Summary}}                                         \\ \hline
\multicolumn{1}{|l|}{\textbf{Populations}}    & \multicolumn{1}{X|}{Four: excitatory, inhibitory, excitatory external input, inhibitory external input}                  \\ \hline
\multicolumn{1}{|l|}{\textbf{Topology}}       & \multicolumn{1}{X|}{-}                                                                                                   \\ \hline
\multicolumn{1}{|l|}{\textbf{Connectivity}}   & \multicolumn{1}{X|}{Random convergent connections with probability $P$ and fixed in-degree of $C_E = PN_E$ and $C_I=PN_I$} \\ \hline
\multicolumn{1}{|l|}{\textbf{Neuron model}}   & \multicolumn{1}{X|}{Leaky integrate-and-fire, fixed voltage threshold, fixed absolute refractory time (voltage clamp)}   \\ \hline
\multicolumn{1}{|l|}{\textbf{Channel models}} & \multicolumn{1}{X|}{-}                                                                                                   \\ \hline
\multicolumn{1}{|l|}{\textbf{Synapse model}}  & \multicolumn{1}{X|}{$\delta$-current inputs}                                                                             \\ \hline
\multicolumn{1}{|l|}{\textbf{Plasticity}}     & \multicolumn{1}{X|}{-}                                                                                                   \\ \hline
\multicolumn{1}{|l|}{\textbf{Input}}          & \multicolumn{1}{X|}{Independent fixed-rate Poisson spike trains to all neurons}                                          \\ \hline
\multicolumn{1}{|l|}{\textbf{Measurements}}   & \multicolumn{1}{X|}{Spike timing, population firing rate.}                                                               \\ \hline
\end{tabularx}

\begin{tabularx}{\textwidth}{llX}
\hline
\rowcolor[HTML]{333333} 
{\color[HTML]{FFFFFF} B}            & \multicolumn{2}{c}{{\color[HTML]{FFFFFF} Populations}} \\ \hline
\multicolumn{1}{|l|}{\textbf{Name}} & \multicolumn{1}{l|}{\textbf{Elements}} & \multicolumn{1}{X|}{\textbf{Size}}    \\ \hline
\multicolumn{1}{|l|}{$E$}           & \multicolumn{1}{l|}{Iaf neuron}        & \multicolumn{1}{X|}{$N_E $}     \\ \hline
\multicolumn{1}{|l|}{$I$}           & \multicolumn{1}{l|}{Iaf neuron}        & \multicolumn{1}{X|}{$N_I = \gamma N_E = 0.25N_E$}            \\ \hline
\multicolumn{1}{|l|}{$E_{\text{ext}}$}     & \multicolumn{1}{l|}{Poisson generator} & \multicolumn{1}{X|}{$C_E(N_E + N_I)$} \\ \hline
\multicolumn{1}{|l|}{$I_{\text{ext}}$}     & \multicolumn{1}{l|}{Poisson generator} & \multicolumn{1}{X|}{$C_I(N_E + N_I)$} \\ \hline
\end{tabularx}

\begin{tabularx}{\textwidth}{lllX}
\hline
\rowcolor[HTML]{333333} 
{\color[HTML]{FFFFFF} C}            & \multicolumn{3}{c}{{\color[HTML]{FFFFFF} Connectivity}}                                                                           \\ \hline
\multicolumn{1}{|l|}{\textbf{Name}} & \multicolumn{1}{l|}{\textbf{Source}} & \multicolumn{1}{l|}{\textbf{Target}} & \multicolumn{1}{l|}{\textbf{Pattern}}                                       \\ \hline
\multicolumn{1}{|l|}{$EE$}          & \multicolumn{1}{l|}{$E$}               & \multicolumn{1}{l|}{$E$}               & \multicolumn{1}{X|}{Random convergent $C_E \to 1$, weight $J$, delay $D$}   \\ \hline
\multicolumn{1}{|l|}{$IE$}          & \multicolumn{1}{l|}{$I$}               & \multicolumn{1}{l|}{$E$}               & \multicolumn{1}{X|}{Random convergent $C_E \to 1$, weight $J$, delay $D$}   \\ \hline
\multicolumn{1}{|l|}{$EI$}          & \multicolumn{1}{l|}{$E$}               & \multicolumn{1}{l|}{$I$}               & \multicolumn{1}{X|}{Random convergent $C_I \to 1$, weight $-gJ$, delay $D$} \\ \hline
\multicolumn{1}{|l|}{$II$}          & \multicolumn{1}{l|}{$I$}               & \multicolumn{1}{l|}{$I$}               & \multicolumn{1}{X|}{Random convergent $C_I \to 1$, weight $-gJ$, delay $D$} \\ \hline
\multicolumn{1}{|l|}{$\text{Ext}_{\text{exc}}$}   & \multicolumn{1}{l|}{$E_{\text{ext}}$}       & \multicolumn{1}{l|}{$E \cup I$}      & \multicolumn{1}{X|}{Non-overlapping $C_E \to 1$, weight $J_{\text{ext}}$, delay $D$} \\ \hline
\multicolumn{1}{|l|}{$\text{Ext}_{\text{inh}}$}   & \multicolumn{1}{l|}{$I_{\text{ext}}$}       & \multicolumn{1}{l|}{$E \cup I$}      & \multicolumn{1}{X|}{Non-overlapping $C_I \to 1$, weight $J_{\text{ext}}$, delay $D$} \\ \hline
\end{tabularx}
\fixedit{\tiny K: What does ``random convergent" mean? JB: "Random convergent" means that each neurons has a fixed number of pre-syaptic neurons chosen randomly from the other neurons in the network. It is opposed to "random divergent" where each neurons has a fixed number of post-synaptic neurons. This is covered in \mycite{Nordlie2009} if we want to had a reference. G: Since this table says it refers to the formalism of \mycite{Nordlie2009}, I'm fine with not saying anything more about what terms mean.} 
\begin{tabularx}{\textwidth}{lX}
\hline
\rowcolor[HTML]{333333} 
{\color[HTML]{FFFFFF} D} & \multicolumn{1}{c}{{\color[HTML]{FFFFFF} Neuron and Synapse Model}}   \\ \hline
\multicolumn{1}{|l|}{\textbf{Name}} & \multicolumn{1}{X|}{- (Brian2)} \\ \hline
\multicolumn{1}{|l|}{\textbf{Type}} & \multicolumn{1}{X|}{leaky integrate-and-fire, $\delta$-synapse} \\ \hline
\multicolumn{1}{|l|}{} & \multicolumn{1}{X|}{$\tau_m \dot{V}_m(t) = -V_m(t) + RI(t)$ \hspace{0.5cm} if not refractory ($t > t^{\ast} +\tau_{rp}$)} \\
\multicolumn{1}{|l|}{} & \multicolumn{1}{X|}{$V_m(t) = V_r$ \hspace{3.cm} while refractory ($t^{\ast} < t \leq t^{\ast} +\tau_{rp}$)} \\
\multicolumn{1}{|l|}{\multirow{-3}{*}{\textbf{Membrane potential}}} & \multicolumn{1}{X|}{$I(t) = \frac{\tau_m}{R}\sum_{\tilde{t}}\omega\delta(t-(\tilde{t}+D))$} \\ \hline
\multicolumn{1}{|l|}{} & \multicolumn{1}{X|}{if $V(t-)<\theta \wedge V(t+)\geq\theta$} \\
\multicolumn{1}{|l|}{} & \multicolumn{1}{X|}{1. set $t^{\ast} = t$} \\
\multicolumn{1}{|l|}{\multirow{-3}{*}{\textbf{Spiking}}} & \multicolumn{1}{X|}{2. emit spike with time-stamp $t^{\ast}$} \\ \hline
\end{tabularx}

\begin{tabularx}{\textwidth}{|l|X|}
\rowcolor[HTML]{333333} 
{\color[HTML]{FFFFFF} E} & \multicolumn{1}{c}{\color[HTML]{FFFFFF} Input} \\
\multicolumn{1}{|l|}{\textbf{Type}} & \multicolumn{1}{X|}{\textbf{Description}} \\ \hline
Poisson generators & \multicolumn{1}{X|}{Fixed rate $\nu$, $C_E+C_I$ generators per neuron, each generator projects to one neuron:}                \\
                   & \multicolumn{1}{X|}{\hspace{1cm} if excitation dominates ($g\leq4$): $E_{\text{ext}}$ rate = $\nu_{\text{ext}}$, $I_{\text{ext}}$ rate = 0}                      \\
                   & \multicolumn{1}{X|}{\hspace{1cm} if inhibition dominates ($g>4$): $E_{\text{ext}}$ rate = $\nu_{\text{ext}} + \nu_{\text{bal}} $, $I_{\text{ext}}$ rate = $\nu_{\text{bal}}/g$} \\ \hline
\end{tabularx}

\centering
\caption{Description of the neuronal network following the formalism of~\mycite{Nordlie2009}, part 1/2.}
\label{tab:brunel_model_1}
\end{table}

\begin{table}[htbp]
\begin{tabularx}{\textwidth}{lX}
\rowcolor[HTML]{333333} 
{\color[HTML]{FFFFFF} F} & \multicolumn{1}{c}{\color[HTML]{FFFFFF} Measurement} \\ \hline
\multicolumn{2}{|X|}{Spiking times of all neurons, population firing rate} \\ \hline
\end{tabularx}
  
\begin{tabularx}{\textwidth}{lXllll}
\rowcolor[HTML]{333333} 
{\color[HTML]{FFFFFF} G}              & \multicolumn{5}{c}{{\color[HTML]{FFFFFF} Network parameters}}             \\ \hline
\multicolumn{2}{|l|}{}                                     & \multicolumn{4}{c|}{\textbf{Network configuration}}                          \\
\multicolumn{2}{|X|}{\textbf{Parameters}}                  & Common       & ver. 1       & ver. 2      & \multicolumn{1}{l|}{ver. 3}      \\ \hline
\multicolumn{2}{|X|}{Number of excitatory neurons $N_E$}   & 2000         & -            & -           & \multicolumn{1}{l|}{-}           \\
\multicolumn{2}{|X|}{Number of inhibitory neurons $N_I$}   & 500          & -            & -           & \multicolumn{1}{l|}{-}           \\
\multicolumn{2}{|X|}{Excitatory synapses per neuron $C_E$} & -            & 200          & 800         & \multicolumn{1}{l|}{800}         \\
\multicolumn{2}{|X|}{Inhibitory synapses per neuron $C_I$} & -            & 50           & 200         & \multicolumn{1}{l|}{200}         \\ \hline
\end{tabularx}

\begin{tabularx}{\textwidth}{lXllll}
\rowcolor[HTML]{333333} 
{\color[HTML]{FFFFFF} H}                                 & \multicolumn{5}{c}{{\color[HTML]{FFFFFF} Neurons parameters}}                               \\ \hline
\multicolumn{2}{|l|}{}                                                                          & \multicolumn{4}{c|}{\textbf{Network configuration}}                          \\
\multicolumn{2}{|X|}{\textbf{Parameters}}                                                       & Common       & ver. 1       & ver. 2      & \multicolumn{1}{l|}{ver. 3}      \\ \hline
\multicolumn{2}{|X|}{Membrane time constant $\tau_{m}/ms$}                                      & 20           & -            & -           & \multicolumn{1}{l|}{-}           \\
\multicolumn{2}{|X|}{Refractory period $\tau_{rp}/ms$}                                          & 2            & -            & -           & \multicolumn{1}{l|}{-}           \\
\multicolumn{2}{|X|}{Firing threshold $V_{\theta}/mV$}                                          & 20           & -            & -           & \multicolumn{1}{l|}{-}           \\
\multicolumn{2}{|X|}{Resting potential $V_{\text{rest}}/mV$}                                           & 0            & -            & -           & \multicolumn{1}{l|}{-}           \\
\multicolumn{2}{|X|}{Reset potential $V_r/mV$}                                                  & 10           & -            & -           & \multicolumn{1}{l|}{-}           \\
\multicolumn{2}{|X|}{Post-synaptic potential (PSP) amplitude from reccurent connections $J/mV$}  & -            & 0.5          & 1.0         & \multicolumn{1}{l|}{1.0}         \\
\multicolumn{2}{|X|}{PSP amplitude from external connections $J_{\text{ext}}/mV$:}                     &              &              &             & \multicolumn{1}{l|}{}            \\
\multicolumn{2}{|X|}{\hspace{2cm} if excitation dominates ($g \leq 4$)}                                    & -            & 0.5          & 1.0         & \multicolumn{1}{l|}{1.0}         \\
\multicolumn{2}{|X|}{\hspace{2cm} if inhibition dominates ($g > 4$)}                                    & -            & 0.1          & 0.2         & \multicolumn{1}{l|}{0.2}         \\
\multicolumn{2}{|X|}{Synaptic delay $D/ms$}                                                     & -            & 1.5          & 1.5         & \multicolumn{1}{l|}{3}           \\ \hline
\end{tabularx}

\centering
\caption{Description of the neuronal network following \mycite{Nordlie2009}, part 2/2.}
\label{tab:brunel_model_2}
\end{table}

\subsection{Topological framework} 
In \emph{algebraic topology}, a well-established field of mathematics, one studies \emph{topological spaces} by turning them into well-behaved \textit{algebraic invariants} and deducing properties of the spaces from those of the algebraic objects. We shall not define any of these concepts precisely here, but will instead give relevant examples of both. See for example~\mycite{hatcher} for an introductory textbook that includes all the details with full precision.

A space will in our context mean a kind of object that is built from certain geometric pieces by specific rules that reflect the data of the dynamics (or structure) of systems of neurons. These objects are ``high-dimensional'' in the sense that they can be thought of as inhabiting $\RR^n$ for (possibly) very large $n$, but the way that they do so is of no relevance to the algebraic invariants we employ here. Moreover, the Euclidean coordinatization of the space in general contains no information about the underlying system, so it is therefore to our benefit that the topological methods ignore it.

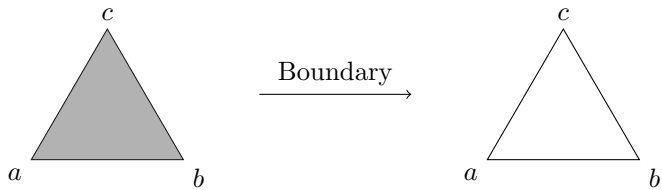
\begin{figure}[htbp]
  \centering
  \begin{tikzpicture}
    \begin{scope}
      \coordinate (a) at (0,0);
      \coordinate (b) at (0:2);
      \coordinate (c) at (60:2);
      \draw[twosimp] (a) -- (b) -- (c) -- cycle;
      \node[anchor=north east] () at (a.south west) {$a$};
      \node[anchor=north west] () at (b.south east) {$b$};
      \node[anchor=south] () at (c.south) {$c$};
      \coordinate (tmp) at ($(a)!0.5!(b)$);
    \end{scope}
    \draw[->] ($(tmp)!0.5!(c) + (2,0)$) -- node[midway,auto] {Boundary} +(2,0);
    \begin{scope}[xshift=6cm]
      \coordinate (a) at (0,0);
      \coordinate (b) at (0:2);
      \coordinate (c) at (60:2);
      \draw (a) -- (b) -- (c) -- cycle;
      \node[anchor=north east] () at (a.south west) {$a$};
      \node[anchor=north west] () at (b.south east) {$b$};
      \node[anchor=south] () at (c.south) {$c$};
    \end{scope}
    \end{tikzpicture}
  \caption{Left: A simplicial complex consisting of a single $2$-simplex ($(a,b,c)$), three $1$-simplices ($(a,b)$, $(b,c)$ and $(a,c)$) and three $0$-simplices ($(a)$, $(b)$ and $(c)$). Right: The boundary of $(a,b,c)$ consists of the three $1$-simplices and three $0$-simplices that one would expect geometrically, giving geometric meaning to a purely combinatorially defined concept.} \label{fig:simplicial}
\end{figure}

A \emph{simplicial complex} can be thought of as a higher-dimensional analog of a graph, with its constituent building blocks referred to as \emph{simplices}. In addition to comprising vertices ($0$-dimensional pieces, or $0$-simplices) and edges ($1$-dimensional pieces, $1$-simplices), a simplicial complex may have filled triangles ($2$-simplices), filled tetrahedra ($3$-simplices), and so on. Just as for graphs, the way these simplices fit together is subject to some rules. First, a $p$-simplex is identified uniquely by a set of $p+1$ vertices. We technically impose a global ordering on all the vertices in the complex and write simplices as tuples respecting this ordering. Second, if a $p$-simplex $\sigma = (v_0, v_1, \dotsc, v_p)$ is present in a simplicial complex $K$, then its $p+1$ \emph{boundary} $(p-1)$-simplices
\begin{equation*}
  (v_1,v_2,v_3,\dotsc, v_p), (v_0, v_2, v_3,\dotsc, v_p), (v_0, v_1,v_2, \dotsc, v_p), \dotsc, (v_0,v_1,v_2,\dotsc, v_{p-1})
\end{equation*}
are all also required to be present in $K$. Note that the definition of a boundary and the associated rule are entirely combinatorial, although they do have a geometric interpretation if we think of simplices as geometric objects, as illustrated in Figure~\ref{fig:simplicial}.

Simplicial complexes can encode data and domain-specific knowledge in different ways depending on the availability of structure and the type of information that is sought. In this work, we take as input a choice of spike train similarity and a spike train for each of $n$ neurons considered. From this, we build a simplicial complex as follows. 

\begin{figure}[htbp]
  \begin{tikzpicture}[scale=0.8]
    \begin{scope}[xshift=-0.2cm]
      \coordinate (a) at (0,0);
      \coordinate (b) at (4,0);
      \coordinate (c) at (4,4);
      \coordinate (d) at (0,4);
      \draw (a) -- node[midway,above] {$\beta$} (b);
      \draw (b) -- node[midway,left] {$\alpha$} (c);
      \draw (c) -- node[midway,below] {$\beta$}(d);
      \draw (d) -- node[midway,right] {$\alpha$} (a);
      \draw (a) -- node[midway,auto] {$\gamma$} (c);
    \end{scope}
    \begin{scope}[xshift = 5cm, yshift = 3.5cm]
      \coordinate (a) at (0,0);
      \coordinate (b) at (2,0);
      \coordinate (c) at (2,2);
      \coordinate (d) at (0,2);
      \draw[fill] (a) circle (2pt);
      \draw[fill] (b) circle (2pt);
      \draw[fill] (c) circle (2pt);
      \draw[fill] (d) circle (2pt);
      \node () at (2.5, 1) {$\subset$};
      \node () at (1, -0.5) {Threshold $0$};
    \end{scope}
    \begin{scope}[xshift = 8cm, yshift = 3.5cm]
      \coordinate (a) at (0,0);
      \coordinate (b) at (2,0);
      \coordinate (c) at (2,2);
      \coordinate (d) at (0,2);
      \draw (a) -- (d);
      \draw (b) -- (c);
      \node () at (2.5, 1) {$\subset$};
      \node () at (1, -0.5) {Threshold $\alpha$};
    \end{scope}
    \begin{scope}[xshift = 11cm, yshift = 3.5cm]
      \coordinate (a) at (0,0);
      \coordinate (b) at (2,0);
      \coordinate (c) at (2,2);
      \coordinate (d) at (0,2);
      \draw (a) -- (d);
      \draw (b) -- (c);
      \draw (c) -- (d);
      \draw (a) -- (b);
      \node () at (2.5, 1) {$\subset$};
      \node () at (1, -0.5) {Threshold $\beta$};
    \end{scope}
    \begin{scope}[xshift = 14cm, yshift = 3.5cm]
      \coordinate (a) at (0,0);
      \coordinate (b) at (2,0);
      \coordinate (c) at (2,2);
      \coordinate (d) at (0,2);
      \draw[twosimp] (a) -- (b) -- (c) -- cycle;
      \draw[twosimp] (a) -- (c) -- (d) -- cycle;
      \node () at (1, -0.5) {Threshold $\gamma$};
    \end{scope}
    
    \begin{scope}[xshift=5cm, yshift=-1.5cm]
      \draw[->, ultra thin] (0.5,0) -- (11,0);
      \draw[->, ultra thin] (0.5,0) -- (0.5,3);
      \node (z) at (1, -0.3) {$0$}; 
      \node (a) at (4, -0.3) {$\alpha$};
      \node (b) at (7, -0.3) {$\beta$};
      \node (c) at (10, -0.3) {$\gamma$};
      \draw[ultra thin] (1, -0.1) -- (1, 0.1);
      \draw[ultra thin] (4, -0.1) -- (4, 0.1);
      \draw[ultra thin] (7, -0.1) -- (7, 0.1);
      \draw[ultra thin] (10, -0.1) -- (10, 0.1);
      
      \draw[ultra thin] (0.4, 0) -- (0.6, 0);
      \draw[ultra thin] (0.4, 0.5) -- (0.6, 0.5);
      \draw[ultra thin] (0.4, 1) -- (0.6, 1);
      \draw[ultra thin] (0.4, 1.5) -- (0.6, 1.5);
      \draw[ultra thin] (0.4, 2) -- (0.6, 2);

      \node () at (0.2, 0) {$0$};
      \node () at (0.2, 0.5) {$1$};
      \node () at (0.2, 1) {$2$};
      \node () at (0.2, 1.5) {$3$};
      \node () at (0.2, 2) {$4$};
      \node[rotate=90] () at (-0.3, 1) {Betti number};
      \node () at (5.5, -0.8) {Filtration threshold (spike train dissimilarity)};

      \node[anchor=east] (label0) at (11, 2.5) {Dimension $0$};
      \node[anchor=east] (label1) at (11, 2) {Dimension $1$};
      \draw (label0) -- +(-2,0);
      \draw[line width=3pt,dashed] (label1) -- +(-2,0);
      
      \draw (0.5,2) -- (1,2) -- (4,2) -- (4,1) -- (7,1) -- (7,0.5) -- (10, 0.5) -- (10.5,0.5);
      \draw[line width=3pt,dashed] (0.5,0) -- (7,0) -- (7,0.5) -- (10,0.5) -- (10, 0) -- (10.5,0);
    \end{scope}
  \end{tikzpicture}
  \caption{Left: A weighted graph $G$ on four vertices/neurons. Assume that $0<\alpha<\beta<\gamma$. Top: The filtered simplicial complex $K$ built from $G$. The $0$-simplices are drawn in a different way at threshold $0$ to make them more visible. Bottom: The Betti curves of $K$ in dimensions $0$ and $1$.} \label{fig:betti}
\end{figure}
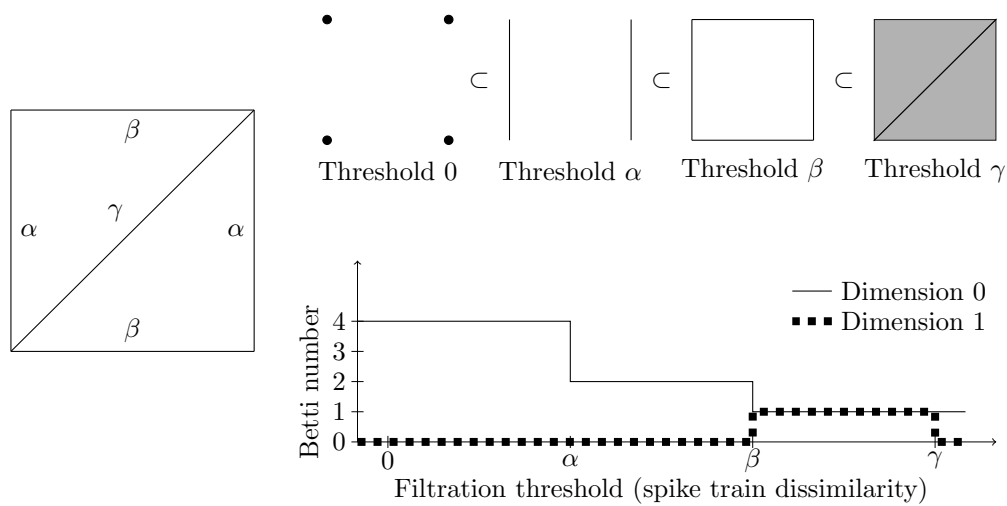

Let $G$ be the complete graph on $n$ vertices (representing the neurons). The edge between two neurons $i$ and $j$ is assigned a weight $w(i,j)$ equal to the dissimilarity of the corresponding spike trains. A simplicial complex $K$ is then formed by adding in every possible $2$-simplex. As a space, this is not very interesting, as there is simply a filled triangle between every triple of neurons (one says that the space is contractible). The crucial part is that each $2$-simplex is given a weight equal to the maximum of the weights given to its boundary edges, i.e.,
\begin{equation*}
  w(i, j, k) = \max\{w(i, j), w(i, k), w(j, k)\}.
\end{equation*}
We then consider a \emph{filtration} of $K$, enabling us to study a sequence of thresholded versions of $K$. At the start of the filtration, the filtration consists only of the vertices of $K$. Then, as the threshold increases, $1$- and $2$-simplices from $K$ appear if their weight is below the threshold, so as to include into the filtration pieces stemming from ever more dissimilar spike trains. See Figure~\ref{fig:betti} for an illustration.

The construction above is applicable to simplices of dimension higher than $2$, so even though we stop at dimension $2$ in our analysis, the following description employs generic dimensions $p$ in order to simplify notation and give the bigger picture.

A basic algebraic invariant that we track as the dissimilarity threshold increases is the \emph{Betti numbers}, giving rise to \emph{Betti curves} for the filtration as a whole. The Betti numbers of a simplicial complex can be defined formally as the dimensions of the \emph{homology} vector spaces of the complex, as we now sketch. Define $C_p(K)$ to be the collection of formal binary sums of the $p$-simplices in $K$. This makes $C_p(K)$ a vector space over the binary numbers, and allows us to view the boundary as an algebraic operation encoded in a linear map $\partial_p:C_p(K)\to C_{p-1}(K)$ given by
\begin{equation*}
  \partial_p(v_0,v_1,\dotsc,v_p) = (v_1,v_2,\dotsc, v_p) + (v_0, v_2, \dotsc, v_p) + (v_0, v_1, \dotsc, v_p) + \dotsm + (v_0,v_1,\dotsc, v_{p-1}).
\end{equation*}
One checks that the application of two consecutive boundary maps always gives zero, i.e., that $\partial_{p-1}(\partial_p(\sigma)) = 0$ for every $p$-simplex $\sigma$. This algebraic statement reflects the geometric fact that the boundary of a boundary is empty. A general collection of $p$-simplices --- a sum in $C_p(K)$ --- that has zero boundary is called a \emph{$p$-cycle}. Figure~\ref{fig:betti} shows several examples.

It turns out that $p$-cycles that are not the boundary of collections of $(p+1)$-simplices correspond geometrically to holes ($p>0$) or connected components ($p=0$) in the simplicial complex. \emph{Persistent homology}, a widely employed construction in topological data analysis, tracks such holes/components as they appear and disappear across a filtration. The record of the ``life'' and ``death'' of such \emph{topological features} provides valuable information about the filtration and thus about the underlying data. We do not use all of the data recorded in persistent homology, but instead just keep track of the \emph{number} of holes in each dimension as a function of the filtration\myfootnote{Reduction to Betti curves}{The reason for reducing the data of persistent homology to Betti curves is in part that the algebraic invariant produced by the former lacks many desirable properties.}. These integer-valued functions, called \emph{Betti curves}, are the features we use for machine learning. An example of a Betti curve is given in Figure~\ref{fig:betti}. We emphasize that the features captured by persistent homology may be much more global in nature than in this small example. As an example, the reader is invited to build a torus as a simplicial complex, filter it from one side to another, and observe what features are captured by persistent homology in dimensions $0$, $1$, and $2$.

\subsection{Machine learning}
Before doing any machine learning, the features selected were standardized. If a feature was not computable because there were no corresponding Betti curves, its value was set to $0$.

For each version of the network, four training sets were formed, one containing $90\%$ of the samples from a specific network version and a fourth one containing $90\%$ of all the samples, stratified so that its distribution of the samples was representative of all the samples. One classifier per training set was trained and tested against four test sets: one for each network version, using the valid samples not in the training set, and the fourth one containing all the valid samples not used during training. 

Support Vector Machine methods \mycite{Cortes1995} using a radial basis function kernel with $L^2$-regularization were applied to classify the samples into the four different regimes (we suspect that a linear classifier would also suffice). The multi-class classification was achieved by training four sub-classifiers with a one-vs-rest decision function. The regularization parameter was found by accuracy optimization thanks to 10-fold cross-validation.

The performance of the classifiers was assessed using an accuracy score.

\subsection{Mutual Information} 
\begin{figure}[htbp]
  \centering
  \includegraphics[width=0.5\textwidth]{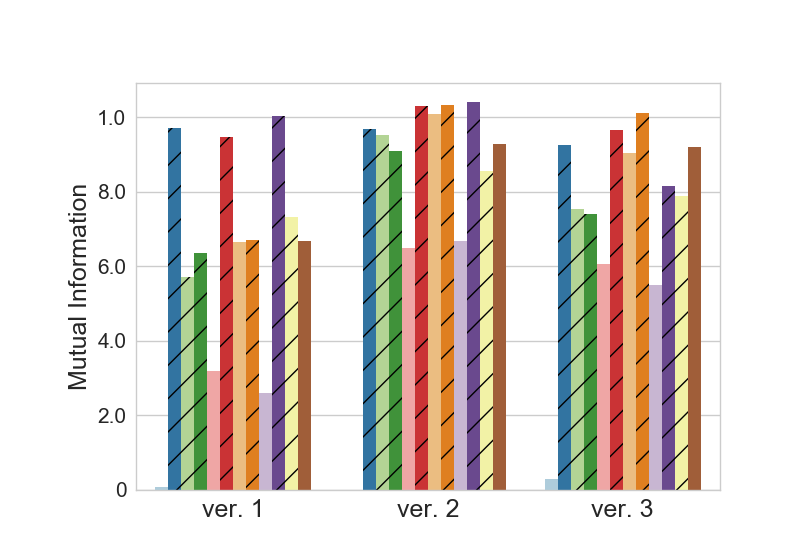}
  \includegraphics[width=0.15\textwidth]{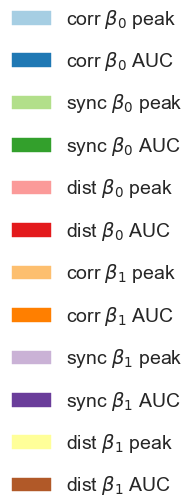}
    \caption[]
  {\small Mutual information between each feature and the label for the training sets obtained from the simulations of the different network versions. Features that were selected by visual inspection are represented with a hashed bar. The \emph{peak} and \emph{AUC} labels designate the peak amplitude and the area under the Betti curve features, respectively.}
  \label{fig:MI_2}
\end{figure}

Earlier we mentioned that mutual information between the features we selected by visual inspection and the regime labels was relatively high, suggesting that one could use the mutual information score to automatically select features when visual inspection would be time consuming or violate a need for automation or independence from human input. The mutual information between each feature and the labels for the three datasets is presented in Figure \ref{fig:MI_2}, where one can observe that some features, such as the area under the Betti-$0$ curve for correlation and SPIKE-distance, have a consistent mutual information score across the three datasets.  This suggests that they are important features that allow the classifier to correctly sample from other datasets. Moreover, the area under the curve (AUC) features tend to have a higher score than the peak amplitude of the Betti curve. This is perhaps natural since the former includes information from all of the filtration, while the latter includes only one single aspect of it.


\section*{Code}
The code for Brunel network simulation, preprocessing before persistent homology computations, preprocessing before machine learning, and for the machine learning itself, is available at \url{https://github.com/JBBardin/Brunel_AlgebraicTopology}. Be aware that the code lacks documentation and is provided as-is for reasons of transparency and reproducibility. The persistent homology computations were performed using Bauer's \textit{Ripser}~\mycite{Bauer2016a}, which is available at \url{https://github.com/Ripser/ripser}.

\section*{Acknowledgments} \label{sec:acknowledgements}
We thank M.-O. Gewaltig for his insightful discussion about the neural network model used.

We are grateful to the Blue Brain Project~\mycite{bbp} for allowing us to use for this project their computational
resources, which are supported by funding from the ETH Domain and hosted at the Swiss National Supercomputing Center (CSCS).

The second author was supported by Swiss National Science Foundation grant number 200021\_172636.

\bibliography{library}
\bibliographystyle{abbrv}

\end{document}